\documentclass[twocolumn,10pt,a4paper,sort&compress,aps,pre,showpacs,superscriptaddress]{revtex4-2}

\usepackage[english]{babel}

\usepackage[caption=false]{subfig}
\addto\captionsenglish{}

\usepackage{overpic}
\usepackage{amsmath,amssymb,amsthm}
\usepackage{commath}
\usepackage{graphicx}
\usepackage{dcolumn}
\usepackage{bm}
\usepackage{hyperref}
\usepackage{etoolbox} 
\usepackage{siunitx}
\usepackage{multirow}
\usepackage[capitalise,nameinlink]{cleveref}
\newcommand{\aref}[1]{\hyperref[#1]{Appendix}}
\usepackage{blkarray}

\usepackage[normalem]{ulem}
\usepackage{comment}
\usepackage[]{xcolor}
\newcommand \bl{\color{black}}

\newcommand \rd{\color{black}}

\newcommand \ke {k^\textrm{e}} 
\newcommand \ko {k^\textrm{o}} 
\newcommand \kn {k^\textrm{ne}} 
\newcommand \kno {k^\textrm{no}} 
\newcommand \kd {k^\textrm{d}} 

\newcommand \erfc {\textrm{erfc}} 

\begin{document}

\title{Odd elastohydrodynamics:
non-reciprocal living material in a viscous fluid}


\author{Kenta Ishimoto}
\email{ishimoto@kurims.kyoto-u.ac.jp}
\affiliation{Research Institute for Mathematical Sciences, Kyoto University, Kyoto 606-8502, Japan}

\author{Cl\'{e}ment Moreau}
\email{cmoreau@kurims.kyoto-u.ac.jp}
\affiliation{Research Institute for Mathematical Sciences, Kyoto University, Kyoto 606-8502, Japan}

\author{Kento Yasuda}
\email{yasudak@kurims.kyoto-u.ac.jp}
\affiliation{Research Institute for Mathematical Sciences, Kyoto University, Kyoto 606-8502, Japan}

\date{\today}

\begin{abstract}

Motility is a fundamental feature of living matter, encompassing single cells and collective behavior. Such living systems are characterized by non-conservativity of energy and a large diversity of spatio-temporal patterns. Thus, fundamental physical principles to formulate their behavior are not yet fully understood. This study explores a violation of Newton's third law in motile active agents, by considering non-reciprocal mechanical interactions known as odd elasticity. By extending the description of odd elasticity to a nonlinear regime, we present a general framework for the swimming dynamics of active elastic materials in low-Reynolds-number fluids, such as wave-like patterns observed in eukaryotic cilia and flagella. We investigate the non-local interactions within a swimmer using generalized material elasticity and apply these concepts to biological flagellar motion. Through simple solvable models and the analysis of {\it Chlamydomonas} flagella waveforms and experimental data for human sperm, we demonstrate the wide applicability of a non-local and non-reciprocal description of internal interactions within living materials in viscous fluids, offering a unified framework for active and living matter physics.
\end{abstract}

\maketitle

\section{Introduction}
Motility is one of the main features of living matter, from a single cell to a swarm of birds or a human crowd \cite{bernheim2018living, das2020introduction, corbetta2023physics}. In the last few decades, the dynamics of motile active agents, both individual and collective behavior, have been intensively studied, giving rise to a rapidly expanding research field in physics bridging non-equilibrium statistical physics, biophysics, and continuum mechanics, now known as active matter and living matter physics.
A crucial feature of these systems is that inner activity units convert energy into mechanical forces. In turn, Newton's third law may be violated when we regard it as an open system, with its mechanical energy being injected from microscopic active units. Therefore, the mechanical interactions between the units can be non-reciprocal \cite{shankar2022topological, klapp2023non}.

The concept of reciprocity is also widely used in continuum mechanics. Recently, violation of the Maxwell-Betti reciprocity in elasticity has been discovered in an active system, and termed odd elasticity \cite{scheibner2020odd, tan2022odd, fruchart2023odd}. The elastic matrix in the constitutive {\bl stress}-strain relation is then allowed to contain non-symmetric components, and it generates a self-sustained propagating wave. 
Odd elasticity reflects the non-conservative forces generated by microscopic active units and provides an effective material constitutive relation for active and living matter. This formulation was shown to effectively describe active locomotion as an autonomous system without controlled, tuned actuation \cite{brandenbourger2021limit}. 

Motile agents at the cellular scale are usually immersed in viscous fluids and are self-propelled by their deformation, as seen in swimming microorganisms \cite{elgeti2015physics, lauga2020fluid}. 
The motility of microswimmers, a term used for active agents in a low-Reynolds-number fluid, is, however, only possible when their deformations are non-reciprocal, which is known as the scallop theorem \cite{purcell1977life, ishimoto2012coordinate, lauga2020fluid}.

Recent theoretical studies on the swimming dynamics of odd-elastic materials \cite{yasuda2021odd,ishimoto2022self} revealed the relations between the violation of Maxwell-Betti reciprocity and the non-reciprocal deformation for microswimming around an equilibrium configuration, demonstrating that the swimming velocity is proportional to the magnitude of odd elasticity.

A traveling wave is a typical example of non-reciprocal deformation ubiquitously observed in biological microswimmers. Indeed, many eukaryotic cells use a flexible slender appendage, called a flagellum or cilium, for propulsion by generating a wave. Examples include tail motions of sperm cells and breaststrokes of {\it Chlamydomonas} green algae \cite{gilpin2020multiscale}.
 This evolutionarily conserved filament is actuated by inner molecular motors in coordination, resulting in a periodic traveling wave with a self-organized nature. The flagellar whip-like motion is therefore regarded as a limit cycle oscillator, and the generic form of flagellar swimming is provided by Hopf bifurcation \cite{camalet2000generic}. Recent theoretical 
and numerical studies using elaborate elastohydrodynamic models also found the emergence of the various flagellar waveform patterns via Hopf bifurcation \cite{de2017spontaneous, ling2018instability, chakrabarti2019spontaneous, man2019morphological}. 
Moreover, refinements of videomicroscopy of biological flagella 
have enabled the detailed analyses of waveforms, and found that the flagellar shape dynamics are well described by a noisy limit cycle that reflects internal activity 
\cite{ma2014active, wan2014rhythmicity,  
saggiorato2017human,
ishimoto2017coarse, ishimoto2018human, guasto2020flagellar}.

The self-sustained wave for an odd-elastic material, however, is insufficient to describe the flagellar waveform, because the odd-elastic waves are dissipated rather than sustained by the fluid viscosity, similar to the classical (passive) elastic response in a viscous medium \cite{lin2023onsager}. Hence, nonlinearity is required for an odd-elastic system to exhibit a stable limit cycle \cite{ishimoto2022self}, calling for a more general, nonlinear odd constitutive relation to deal with biological flagellar swimming.  In fact, the importance of nonlinear odd elasticity has been reported as a topical challenge within the field of active matter studies.\cite{fruchart2023odd}
    
The primary aim of this study is therefore to extend the odd-elastic description of microswimmers to a nonlinear regime to deal with stable periodic deformations, as seen in biological flagellar motion. This theory, which we call {\it odd elastohydrodynamics}, therefore provides a unified framework for the study of non-local, non-reciprocal interactions of an elastic material in a viscous fluid. 

Using this generic formulation, we can access the interactions inside an active elastic material, while these are masked by fluid dynamic coupling when observing flagellar motion under a microscope.
To distinguish the non-reciprocal activity from the passive elastic response, we introduce a new concept, the {\it odd-elastic modulus}, as a spatial Fourier transform in an extended space. The real and imaginary parts of this complex function possess proper symmetry and characterize the reciprocal and non-reciprocal interactions, respectively.

The secondary aim of this study is then to apply our theory to biological flagellar swimmers. By examining the odd-elastic modulus based on simple mathematical models and biological experimental data, we show the
wide applicability of a 
non-local and non-reciprocal description of internal interactions within living materials.

The contents of this paper are summarized as follows. In Section \ref{sec:setup}, we provide a setup for the theoretical formulation of odd elastohydrodynamics to describe an active elastic material in a viscous fluid. 
We also {\bl discuss} the connection between Hopf bifurcation and  nonlinear odd elasticity and express the dynamics of a microswimmer undergoing periodic deformation. {\bl In Section \ref{sec:intro_modulus}, we introduce}
the concept of the odd-elastic modulus.

In Sections \ref{sec:odd_modulus} and \ref{sec:flag_modulus}, we apply our theory to understand the inner mechanical interactions that biological flagellar motion exhibit. 
To gain physical intuition regarding non-local, non-reciprocal interactions encoded by nonlinear odd elasticity, we start with simple and solvable models in Section  \ref{sec:odd_modulus}. We also discuss how the odd-elastic modulus captures the inner interactions of these example models.
In Section \ref{sec:flag_modulus}, we numerically investigate the extended bending modulus in model flagellar waveforms for {\it Chlamydomonas}  and sperm cells, together with experimental data. With these, we propose a new continuum description of living soft matter in a viscous fluid by means of nonlinear odd elasticity. The discussion and conclusions are provided in Section \ref{sec:conc}.

{\bl {\bl One of the advantages of the odd-elastic description of activity is the application of the autonomous equations of motion. These} allow us to analyze some general features of microswimming with periodic deformation, including theoretical formulae for the average swimming velocity. In Appendix \ref{sec:swimming}, to complete our general theory of odd elastohydrodynamics, we further extend our framework to encompass fluctuations in shape gaits by internal actuation, following biological observations of a noisy limit cycle in shape space. {\bl Exploiting the autonomous structure of the odd-elastic formulation and} the gauge-field formulation for microswimming, we investigate the effects of internal active noise on swimming velocity. The role of odd elasticity is further discussed in terms of non-equilibrium thermodynamics.
}

\section{Odd elastohydrodynamics of microswimmers}
\label{sec:setup}

\subsection{Shape and deformation of a swimmer}

\begin{figure}[!t]
\begin{center}
\begin{overpic}[width=8.5cm]{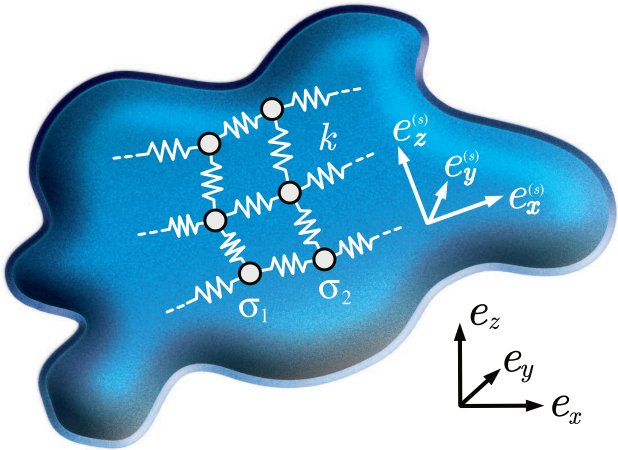}
\end{overpic}
\caption{Schematic of general odd-elastic microswimmer. This example swimmer moves in a three-dimensional space ($d=3$). The position and rotation of the swimmer are represented by the relative motions between the laboratory frame $\{\bm{e}_{x}, \bm{e}_{y}, \bm{e}_{z}\}$ and the swimmer-fixed frame $\{\bm{e}^{(s)}_{x}, \bm{e}^{(s)}_{y}, \bm{e}^{(s)}_{z}\}$. The shape of the swimmer is parameterized by $N$ shape variables $\bm{\sigma}=(\sigma_1, \cdots, \sigma_N)$ In this schematic, we use displacements of material units from equilibrium positions as the shape variables. In typical linear elastic theory, a recovery force is applied that is proportional to the displacement using a spring constant, as indicated by $k$ in this schematic. In the current study, however, we generalize this elastic force to include internal actuation, which is represented by odd elasticity.}
\label{fig:model}
\end{center}
\end{figure}

To describe the motion of a deforming microswimmer in a fluid, we need to specify the position and orientation together with the instantaneous shape of the swimmer. In Fig. \ref{fig:model}, we present a schematic of a general elastic microswimmer.
The rigid body motion is defined by the translation and rotation between the laboratory frame $\{\bm{e}_1,\cdots, \bm{e}_d\}$ and swimmer-fixed frame $\{\bm{e}^{(s)}_{1},\cdots, \bm{e}^{(s)}_{d}\}$. We assume that the swimmer moves in a $d$-dimensional space, where $d=1, 2$, or $3$. $d=1$ indicates linear motion, $d=2$ indicates planar motion, and $d=3$ corresponds to general three-dimensional motion in space. The origin of the swimmer-fixed frame is set to be the swimmer's position and is denoted by $\bm{x}=(x_1, \cdots,x_d)^\textrm{T}$. The number of angular degrees of freedom to specify the orientation in $d$-dimensional space is $d'=d(d-1)/2$. We let $n$ be the number of degrees of freedom for rigid motion, that is, $n=d+d'$, and introduce a $n$-dimensional vector to represent the position and orientation as $\bm{z}_0=(x_1, \cdots, x_d, \theta_1, \cdots, \theta_{d'})^\textrm{T}\in \mathbb{R}^{n}$.

We assume that the shape of the swimmer is parameterized by $N$ shape coordinates as $\bm{\sigma}=(\sigma_1, \dots, \sigma_N)^\textrm{T}\in \mathbb{R}^N$. For the shape coordinates, we employ, for example, displacements from the equilibria of the material units or relative angles between neighboring material units. We will later introduce generalized elastic forces and torques associated with the shape coordinates (see also Fig. \ref{fig:model}).

Let us denote representations in the swimmer-fixed coordinates by superscript ${(s)}$ and 
introduce the extended coordinates vector $\bm{z}=( \bm{z}^{(s)}_0; \bm{\sigma}) \in \mathbb{R}^{n+N}$ and its associated velocity vector, $\dot{\bm{z}}=( \dot{\bm{z}}^{(s)}_0; \dot{\bm{\sigma}})\in \mathbb{R}^{n+N}$, where the semicolon indicates vertical concatenation and the dot symbol indicates a time derivative. The velocity in the swimmer-fixed coordinates $\dot{\bm{z}}^{(s)}_0$ is a physical quantity that is obtained from the force and torque balance equations as explained below. The vector $\bm{z}^{(s)}_0$, computed by integrating $\dot{\bm{z}}^{(s)}_0$, is introduced for later use but does not represent a physical position or orientation when $d=3$ due to the non-commutative nature of the dynamics. 
We set the origin of the shape coordinates, $\bm{\sigma}=\bm{0}$, to be the equilibrium configuration without any internal or external forces.
 
This description includes, in particular, several minimal mathematical models of swimmers. 
For example, Najafi-Golestanian's three-sphere model \cite{najafi2004simple, golestanian2008analytic, yasuda2017elastic} is a swimmer consisting of three spheres connected in a straight line by two rods and moving in one direction by changing the lengths of the rods; thus, the degrees of freedom are $(n, N)=(1, 2)$. Purcell's three-link swimmer \cite{purcell1977life, becker2003self, moreau2019local} is another minimal model, which consists of three rods connected by two hinges to form a snake-like robot, and can swim in a plane by changing the angles of the hinges. The degrees of freedom are therefore $(n, N)=(3, 2)$ for this model. The shape parameters are the lengths of two arms for the three-sphere model and the two relative angles for the three-link swimmer. 

\subsection{Odd-elastohydrodynamic equations}

The dynamics of a three-dimensional self-deforming elastic object in a viscous fluid are well represented by the Stokes equation:
\begin{equation}
\eta\nabla^2\bm{u}=\nabla p
\label{eq:M11}, 
\end{equation}
where the velocity field $\bm{u}$ satisfies the incompressibility condition  $\nabla\cdot\bm{u}=0$. Here, $p$ is the pressure field, and the viscosity $\eta$ is assumed to be constant.
Due to the linearity of the Stokes equations, the hydrodynamic forces and torques conjugate to the extended coordinates, denoted symbolically by $\bm{f}^{\textrm{hyd}}$, and are 
proportional to the time derivative of the extended coordinates. This linear relation is represented by a positive-definite matrix, called a generalized grand resistance matrix ${\bf M}$ \cite{doi2013soft}; hence, $\bm{f}^{\textrm{hyd}}=-{\bf M}\dot{\bm{z}}$.  Due to the negligible inertia, these forces and torques are balanced by internal or external forces and torques, which we denote by $\bm{f}$ and introduce below.

We now define an ``elasticity" matrix (or equivalently an elastic matrix) through a general stress-strain constitutive relation as a function of the shape parameters, ${\bf K}(\bm{\sigma})\in \mathbb{R}^{N\times N}$, to represent all the internal forces and torques, including the internal activity force as well as the ordinary passive elastic response. To be more precise, this generalized elastic matrix is defined by mapping from shape coordinates to internal forces and torques, given by $\bm{f}=-{\bf K}(\bm{\sigma})\bm{ \sigma}$.
This generalized elasticity is reduced to that of an elastic spring when we take the displacement of the material point for the shape coordinates and to that of a torque spring (torsion spring) when we employ the relative angle along a filament as the shape coordinates.
At the equilibrium configuration ($\bm{\sigma}=\bm{0})$, the generalized elastic force vanishes ($\bm{f}=\bm{0}$.) The non-symmetric part may have non-zero values, that is, ${\bf K}\neq {\bf K}^\textrm{T}$; this corresponds to odd elasticity and effectively represents the non-conservative, internal activity of the self-deforming material. If it is linearly odd-elastic, the elastic matrix ${\bf K}$ is a constant matrix, although it is, in general, determined by the instantaneous shape of the object. 
The balance equations for the forces and torque, $\bm{f}^{\textrm{hyd}}+\bm{f}=\bm{0}$, are therefore summarized in the following form \cite{ishimoto2022self}: 
\begin{equation}
-{\bf M}(\bm{\sigma})\dot{\bm{z}}={\bf L}(\bm{\sigma})\bm{z}
\label{eq:M01}.
\end{equation}
Note that the matrix ${\bf M}$ only depends on the spontaneous shape of the swimmer. The right-hand side of Eq. \eqref{eq:M01} represents a general elastic force, including both the passive elastic response and internal actuation, and ${\bf L}$ is a $(n+N)\times (n+N)$ matrix containing the elastic matrix ${\bf K}$ as $L_{n+\alpha, n+\beta}=K_{\alpha\beta}$ with the other components being zero, namely, $L_{ij}=L_{i\alpha}=L_{\alpha j}=0$. Throughout this paper, we use Roman indices such as $i, j=\{ 1,  \dots, n\}$ for the translation and orientation of the object, Greek indices such as $\alpha, \beta =\{1, \dots, N\} $ for the shape coordinates, and the Einstein summation convention for repeated indices.

By inverting the resistance matrix, we can decompose the shape dynamics from the rigid body motion in the form
\begin{equation}
\dot{\bm{z}_0}=-{\bf P}{\bf K}\bm{\sigma}~,~~\dot{\bm{\sigma}}=-{\bf Q}{\bf K}\bm{\sigma}
\label{eq:M01b}.
\end{equation}
The matrices ${\bf P}$ and ${\bf Q}$ are respectively given by $P_{i\alpha}=N_{i, n+\alpha}$ and  $Q_{\alpha\beta}=N_{n+\alpha, n+\beta}$, with ${\bf N}={\bf M}^{-1}$. Note that the second equation \eqref{eq:M01b} provides an autonomous dynamical system in shape space, and the non-symmetric part of the elastic matrix plays the role of an internal actuation to drive the deformation. The first equation determines the translation and rotation of the swimmer and coincides with the equation of the kinematic swimming problem, in which the shape gait is a given function.

\begin{figure}[!t]
\begin{center}
\begin{overpic}
[width=4.5cm]{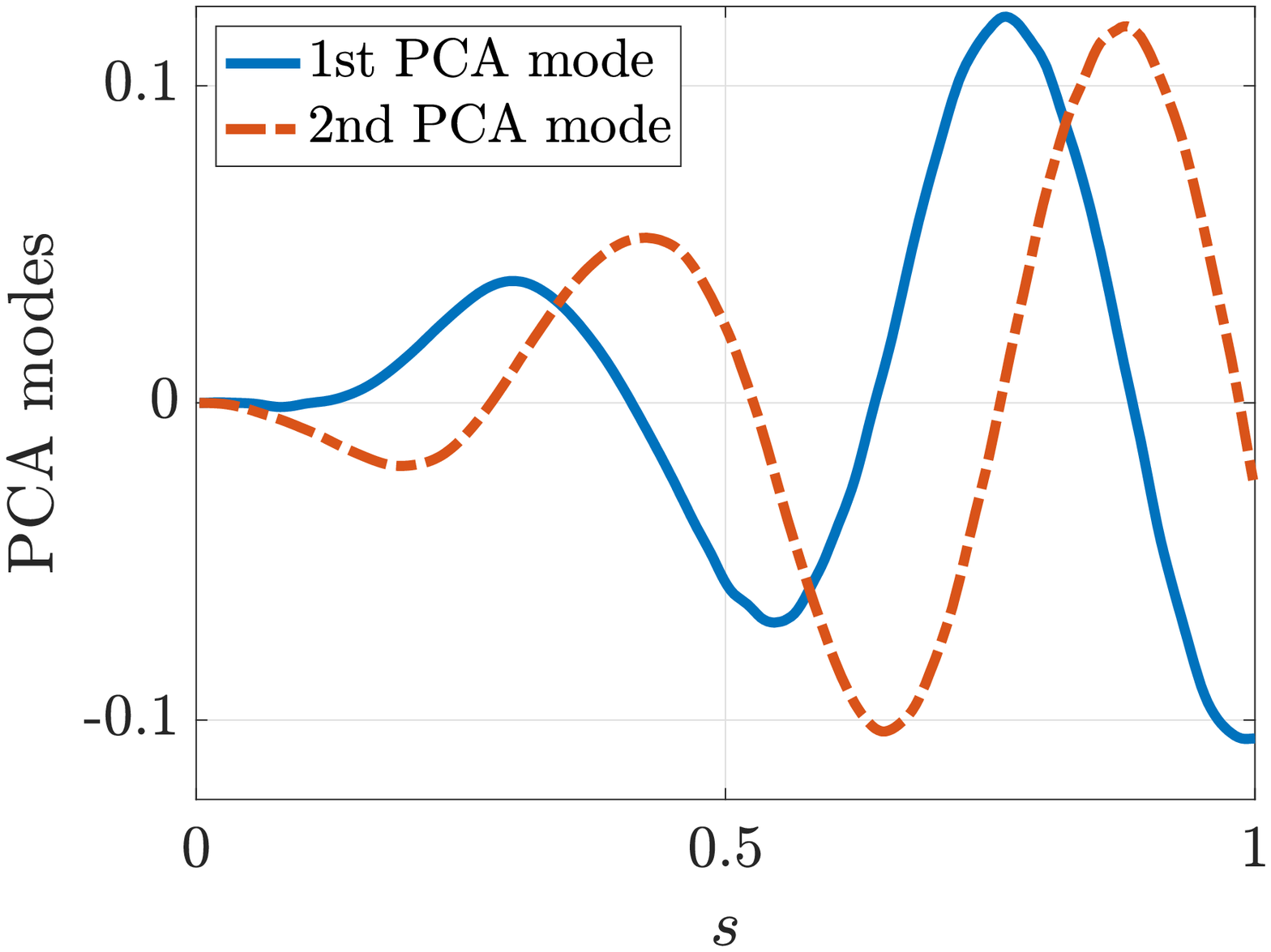}
\put(-4,70){(a)}
\end{overpic}~~~\begin{overpic}
[width=3.5cm]{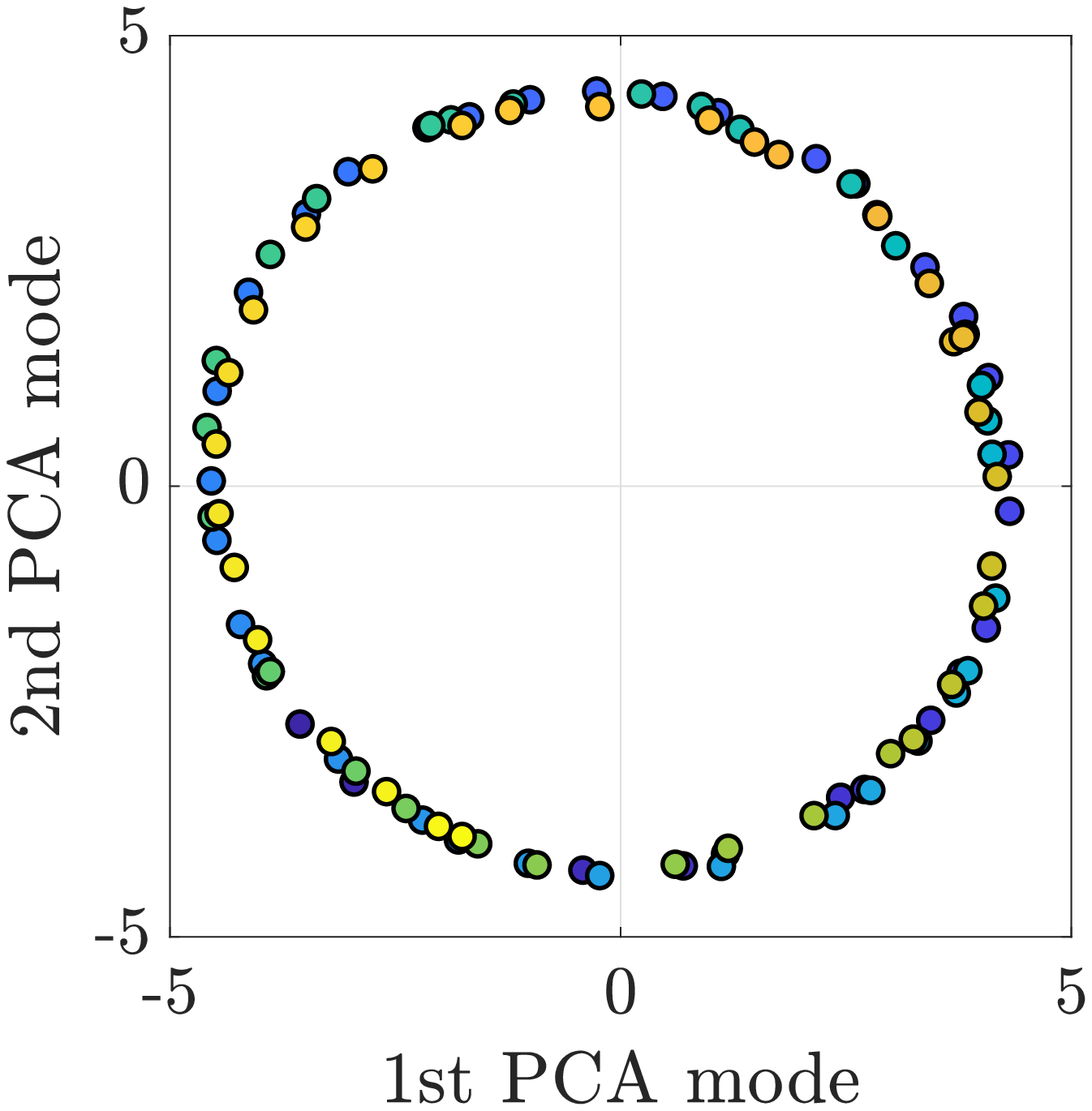}
\put(-4,90){(b)}
\end{overpic}
\\
\vspace*{0.3cm}
~~~~~\begin{overpic}
[width=7.8cm]{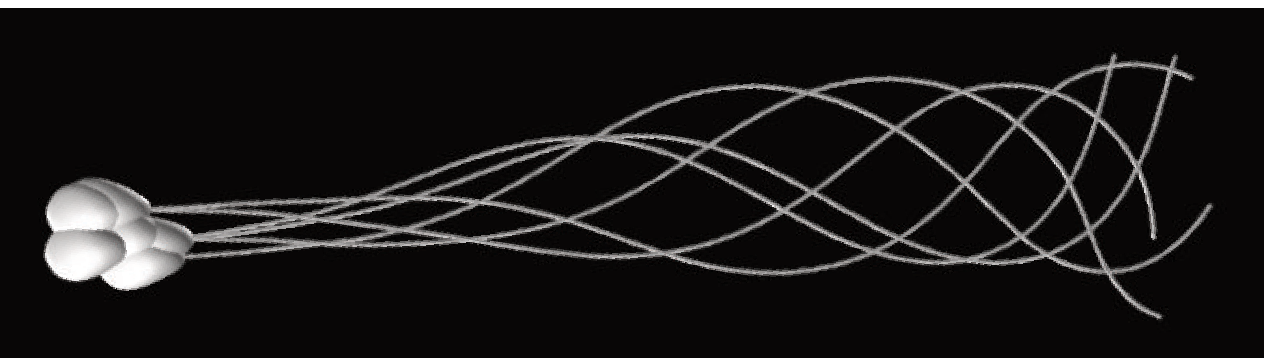}
\put(-9,24){(c)}
\end{overpic}
\caption{Human sperm swimming as example of microswimmer with noisy limit cycle. (a) The lowest two PCA modes were obtained from experimental data. The horizontal axis indicates the normalized arc length along the flagellum from the head-tail junction. (b) Projections of the shape onto the two-dimensional PCA shape space. Each circle indicates the shape at a different time. (c) Superposed snapshots of a swimming human sperm with a time-periodic beat, obtained by a direct numerical simulation of the Stokes equations. The waveform was extracted from the experimental data for swimming human sperm as a limit cycle in the two-dimensional PCA shape space. Figures reproduced from \cite{ishimoto2018human} with permission under the creative commons license (http://creativecommons.org/licenses/by/4.0/).}
\label{fig:hvm}
\end{center}
\end{figure}

\subsection{Periodic swimming by nonlinear odd elasticity }
\label{sec:nonlin_odd}

We now consider a microswimmer undergoing a periodic deformation with a particular focus on flagellar-like filament dynamics. For eukaryotic flagella, internal molecular motors synchronously actuate the elastic filament to generate a periodic waveform. While the emergent waveform is obtained by elastohydrodynamic mechanical coupling, the onset of wave generation from a straight equilibrium configuration is well formulated by a Hopf bifurcation \cite{camalet2000generic, ma2014active, ling2018instability, chakrabarti2019spontaneous}. 

To illustrate the limit cycle behavior in the shape space, we reproduce in Fig. \ref{fig:hvm} the figures on human sperm swimming from Ishimoto et al. \cite{ishimoto2018human}, in which principal component analysis (PCA) was performed to reduce the dimensionality of the flagellar waveform obtained from experimental observations. The authors found that the flagellar waveforms are well represented by noisy limit cycle orbits in the two-dimensional shape space spanned by the lowest PCA modes [Fig. \ref{fig:hvm}(a, b)]. The embedded limit cycle orbit was then extracted and used to analyze the time-periodic swimming dynamics of human sperm [Fig. \ref{fig:hvm}(c)]. 

To derive a generic description for such time-periodic swimming, we employ the normal form of the Hopf bifurcation, which may be written as \cite{guckenheimer2013nonlinear}
\begin{equation}
\frac{dz}{dt}=cz+b|z|^2z,
\label{eq:N11a}
\end{equation}
where $z\in\mathbb{C}$. The parameters $c$ and $b$ are both complex numbers: $c=\lambda+i\omega\in\mathbb{C}$ and $b=\mu+i\xi \in\mathbb{C}$ with real-valued parameters $\lambda, \omega, \mu$, and $\xi$.
Let us introduce the {\it apparent} shape space in which the shape dynamics are described by the normal form \eqref{eq:N11a} and 
denote the shape coordinates in this shape space by $\bm{q}$. The apparent shape coordinates do not always coincide with the shape coordinates $\bm{\sigma}$ used in the stress-strain relation. To distinguish $\bm{\sigma}$ from $\bm{q}$, we now refer to $\bm{\sigma}$ as {\it intrinsic} shape coordinates.  
We then assume the existence of a transformation from the apparent shape coordinates to the intrinsic shape coordinates given by a full-rank matrix ${\bf W} \in \mathbb{R}^{N\times N}$, that is, $\bm{\sigma}={\bf W}\bm{q}$. The matrix ${\bf W}$ may be obtained by PCA and we will later examine detailed construction of the matrix with some examples (Sections \ref{sec:odd_modulus} and \ref{sec:flag_modulus}).

From the normal form of Eq. \eqref{eq:N11a}, the dynamics in the apparent shape space are separated into the limit cycle in the $q_1-q_2$ space and the damping dynamics in the remaining $(N-2)$ dimensions. Let us introduce the {\it apparent} elastic matrix $\hat{{\bf K}} \in\mathbb{R}^{N\times N}$ to distinguish the apparent elasticity from the {\it intrinsic} elasticity ${\bf K}$ in Eq. \eqref{eq:M01b}, and write the dynamics in the form \begin{equation}
\dot{\bm{q}}=-\hat{\bf K}\bm{q}
    ~~ \textrm{with}~~
{\bf \hat{K}}(\bm{q})=
\begin{pmatrix}
{\bf \hat{K}}^{\textrm{LC}} & {\bf O} \\
{\bf O} & {\bf \hat{K}}^{\textrm{d}}
\end{pmatrix}
\label{eq:N10a},
\end{equation}
where the two-dimensional nonlinear elastic matrix ${\bf \hat{K}}^{\textrm{LC}}\in \mathbb{R}^{2\times 2}$ represents the limit cycle in Eq. \eqref{eq:N11a}, and the $(N-2)$-dimensional matrix ${\bf \hat{K}}_{d}\in \mathbb{R}^{(N-2)\times (N-2)}$ in the right-bottom block expresses the stable modes around the Hopf bifurcation. All the eigenvalues of ${\bf \hat{K}}^{\textrm{d}}$ therefore have non-negative real parts. 

After relabeling the parameters $\lambda, \omega, \mu,$ and $\xi$
in Eq. \eqref{eq:N10a} as $\ke, \ko, \kn,$ and $\kno$ with an additional minus sign, we may write the components of ${\bf \hat{K}^{\textrm{LC}}}$ as
\begin{equation}
\hat{K}^{\textrm{LC}}_{\alpha\beta}=(\ke+\kn r^2)\delta_{\alpha\beta}+(\ko+\kno r^2)\epsilon_{\alpha\beta}
\label{eq:M11b}
\end{equation}
for $\alpha, \beta \in\{1, 2\}$. Here,  $\delta_{\alpha\beta}$ is the Kronecker delta, $\epsilon_{\alpha\beta}$ is the two-dimensional Levi-Civita permutation symbol, and $r=(q_1^2+q_2^2)^{1/2}$. 
With these terms, the dynamics in the apparent shape coordinates are translated from the normal form of the Hopf bifurcation into dynamics described by odd-elastic interactions.
The four parameters $\ke, \ko, \kn$, and $\kno$ are
 then interpreted as even linear elasticity, odd linear elasticity, even nonlinear elasticity, and odd nonlinear elasticity, respectively.
 

{\bl 
 In this Section \ref{sec:nonlin_odd}, we introduced the normal form for the limit cycle in shape space, which then couples with hydrodynamics to generate the net displacement, i.e., locomotion or swimming. In this theoretical framework, the swimming dynamics are fully described by an autonomous system. Hence, by integrating this system over the cycle of shape deformation, we may obtain a general formula {\bl for the average swimming velocity for a small-amplitude swimmer \cite{shapere1989geometry, ishimoto2022self}. }}


The position and the orientation in $d$ dimensions are represented by an element of the $d$-dimensional Euclidean group $\textrm{SE}(d)$. We represent these by $\mathcal{R}\in \mathbb{R}^{n\times n}$ and the time evolution is provided by its generator, $\mathcal{A}$, via $\dot{\mathcal{R}}=\mathcal{R}\mathcal{A}$. With the linearity of the Stokes equation, we may rewrite this generator as $\mathcal{A}=\mathcal{A}_\alpha q_\alpha$, and the third-rank tensor $[\mathcal{A}_\alpha]_{ij}=A_{ij\alpha}$ is the connection of the gauge group $\textrm{SE}(d)$.  If the swimmer exhibits a periodic motion with period $T_c$, the displacement and rotation after one beat cycle are obtained by a loop integral in shape space \cite{shapere1989geometry, koiller1996problems, avron2008geometric} as
 \begin{equation}
 \mathcal{R}(T)=\mathcal{R}_0\bar{\textrm{P}} \exp\left[\int_0^{T_c}\mathcal{A}(t)\,dt\right]
 =\mathcal{R}_0\overline{\textrm{P}} \exp\left[\oint \mathcal{A}_\alpha\,dq_\alpha\right]
 \label{eq:M04},
 \end{equation} 
where we write $\mathcal{R}(t=0)=\mathcal{R}_0$ and introduce a path-ordering operator $\overline{\textrm{P}}$. The integral in the last term is performed over a closed loop in shape space. After expanding for a small $\bm{q}$ {\bl up to its quadratic term}, the swimming velocity over one beat cycle is obtained for a small-amplitude deformation by using a fourth-rank tensor $\mathcal{F}$, called the curvature of the gauge field, as
\begin{equation}
\overline{A_{ij}}=\frac{1}{2} F_{ij\alpha\beta}\overline{q_\alpha\dot{q}_{\beta}}
\label{eq:M05},
\end{equation}
where the overline indicates the average over one deformation cycle.
{\bl
Hence, the average swimming velocity is proportional to the areal velocity enclosed by the limit cycle in shape space. Using our odd-elastic representation, Eqs. \eqref{eq:N10a}-\eqref{eq:M11b}, the limit cycle exists only when $\ke<0$. Then, the swimming formula can be computed as 
\begin{equation}
\overline{A_{ij}}=F_{ij12}
\left[\frac{\ko}{2}
\frac{|\ke|}{\kn}+\frac{\kno}{2}\left(\frac{|\ke|}{\kn}\right)^2
\right]
    \label{eq:M05d}.
\end{equation}
The terms on the right-hand side are proportional to the odd-elastic coefficients. This equation generalizes the swimming formula for an odd-elastic swimmer from the linear to the non-linear regime \cite{ishimoto2022self}.}
 
{\bl
 In addition, as observed in many biological systems, the shape gait, or the coefficients for the odd elasticity in our framework, temporally fluctuates.
 Further, these active fluctuations provides an important link to non-equilibrium statistical physics. 
 Therefore, internal noise should be taken into consideration to precisely evaluate locomotion.  Thus, to complete our theory, we thoroughly investigated the impact of noise on the swimming velocity, extending the swimming formula \eqref{eq:M05d} for noisy limit cycles. The detailed calculations can be found in Appendix \ref{sec:swimming}.
 }



\section{Non-local, non-reciprocal interactions and odd-elastic modulus}
\label{sec:intro_modulus}

{\bl In this section, we focus on the interactions between the units of active material undergoing periodic deformation in a viscous fluid. To characterize its non-local, non-reciprocal interactions, we introduce the concept of the odd-elastic modulus.}

By changing the variables from $\bm{q}$ to $\bm{\sigma}$, the intrinsic elastic matrix can be derived from the apparent elastic matrix as
\begin{equation}
{\bf K}={\bf Q}^{-1}{\bf W}{\bf \hat{K}}{\bf W}^{-1}
    \label{eq:I01}.
\end{equation}

As already described, the elastic force or torque including the passive and active elastic response is symbolically given by
\begin{equation}
f_\alpha=-K_{\alpha\beta}\sigma_\beta
    \label{eq:I02a}.
\end{equation}
When considering the intrinsic shape coordinates, we usually chose the displacement from the equilibrium for a material unit and the relative distance or angle between neighboring material units. 

{\bl 
In the example of an odd three-sphere swimmer \cite{yasuda2021odd}, the intrinsic elastic matrix is given by 
\begin{equation}
K_{\alpha\beta}=\ke \delta_{\alpha\beta}+\ko \epsilon_{\alpha\beta}
    \label{eq:I02a1},
\end{equation}
for $\alpha, \beta \in \{1,2\}$. For a Purcell swimmer with odd-elastic hinges \cite{ishimoto2022self}, the same intrinsic elastic matrix is considered, where $f_i$ indicates the torque at the $i$-th hinge and $\sigma_i$ is the $i$-th relative angle between neighboring rods.

\begin{figure}[!t]
\begin{center}
\begin{overpic}[width=\linewidth]
{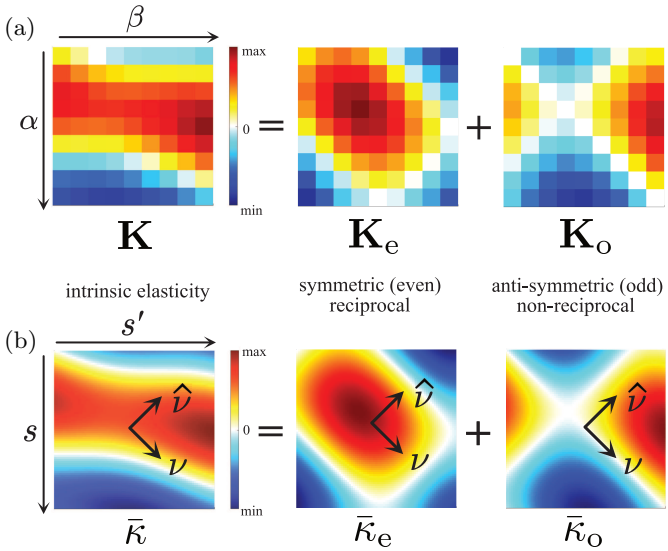}
\put(-3,80){{ (a)}}
\put(-3,32){{ (b)}}
\end{overpic}
\caption{{\bl Schematics of elastic matrix $K_{\alpha\beta}$ and its continuum representation $\bar{\kappa}(s, s')$ with decomposition into symmetric even elasticity (reciprocal interaction) and anti-symmetric odd elasticity (non-reciprocal interaction), (a) ${\bf K}={\bf K}_{\textrm{e}}+{\bf K}_{\textrm{o}}$ and (b) $\bar{\kappa}=\bar{\kappa}_{\textrm{e}}+\bar{\kappa}_{\textrm{o}}$. The two-dimensional Fourier transform is associated with a two-dimensional wave vector $(\nu_{s}, \nu_{s'})$. To characterize the non-reciprocal nature of the interactions encoded in the intrinsic elasticity, we consider the Fourier modes along the diagonal components and in the perpendicular direction, respectively indicated by $\nu$ and $\hat{\nu}$ in the schematics. The odd-elastic modulus, Eq. \eqref{eq:I21b}, is defined by the Fourier modes in the $\hat{\nu}$ direction.}}
\label{fig:mat-schmat}
\end{center}
\end{figure}

We further extend this form into a $N\times N$ matrix representation. A schematic is shown in Fig. \ref{fig:mat-schmat}(a), where the size of the matrix is set to $N=10$, with the colors indicating the values of the matrix components, which are chosen arbitrarily for illustration purposes. The off-diagonal components express non-local interactions. The matrix is then decomposed into its symmetric and anti-symmetric parts, {\bl i.e., ${\bf K}={\bf K}_{\textrm{e}}+{\bf K}_{\textrm{o}}$}. These correspond to the even and odd elastic matrices, respectively, and the anti-symmetric matrix represents the non-reciprocal interactions between the units of material.
}

Let us introduce the Lagrangian coordinate of the material for a point with shape index $\alpha$, {\bl which is denoted by position $\bm{s}_\alpha \in \mathbb{R}^d$. For simplicity and later use, here we focus on a one-dimensional elastic object such as a filament or rod, and assume its length at rest to be $\ell$. For the Lagrangian coordinates, we take an equally-spaced arclength along the material at rest and represent it by $s_\alpha\in [0, \ell]$ with its separation being $\Delta\ell=\ell/N$.} 
{\bl We then rewrite Eq. \eqref{eq:I02a}, representing the force or torque acting on a material point with a Lagrange label $s_\alpha$,  by a non-local interaction represented by a kernel, $\kappa(s_{\alpha}, s_{\beta})$,} as 
\begin{equation}
f(s_\alpha)=-\sum_{\beta}\kappa(s_\alpha, s_\beta)\sigma_j(s_\beta)
    \label{eq:I02a2}.
\end{equation}
{\bl For large $N$, it is also useful to consider a continuum representation.  By dividing both terms in \eqref{eq:I02a2} by the spatial discretization, we obtain
\begin{equation}
  \frac{f(s_\alpha)}{\Delta \ell}=-\sum_{\beta}\frac{\kappa(s_\alpha, s_\beta)}{\Delta \ell}\frac{\sigma_j(s_\beta)}{\Delta \ell}\,\Delta\ell
    \label{eq:I02a3}.  
\end{equation}
Each term in Eq.\eqref{eq:I02a3} represents the density (quantity per unit length): 
$\bar{f}=f/\Delta \ell$ and $\bar{\sigma}=\sigma/\Delta \ell$ are the force or torque per unit length and  the displacement per unit length, respectively, whereas $\bar{\kappa}=\kappa/\Delta \ell$ is interpreted as an extension of the elastic modulus. 

In the continuum representation, as is usually considered, we assume that these densities are well-behaved functions.} Hence, by replacing the summation with an integral, the continuous form in the large-$N$ limit is obtained as
{\bl
\begin{equation}
\bar{f}(s)=-\int_0^\ell \bar{\kappa}(s, s')\bar{\sigma}(s')\,ds'
    \label{eq:I02b},
\end{equation}
}
where the integral is performed over the material.


A non-energy-conserving active material may possess non-symmetric components in the elastic matrix ($K_{\alpha\beta}\neq K_{\beta\alpha}$), and this corresponds to non-reciprocity of non-local elastic interactions:
\begin{equation}
{\bl
\bar{\kappa}(s, s')\neq \bar{\kappa}(s', s)
}
    \label{eq:I03}.
\end{equation}
{\bl As illustrated in Fig. \ref{fig:mat-schmat}(b), we then decompose the kernel function into reciprocal (even) and non-reciprocal (odd) components as $\bar{\kappa}(s,s')=\bar{\kappa}_{\textrm{e}}(s,s')+\bar{\kappa}_{\textrm{o}}(s,s')$.}
Note that the non-reciprocity expressed by Eq. \eqref{eq:I03} refers to the breakdown of the symmetry under the change of two material points, $(s, s')\mapsto(s', s)$, and differs from the reciprocity with respect to the physical coordinates, $(i,j)\mapsto (j,i)$.
The macroscopic odd elasticity often refers to  non-reciprocity in the physical coordinates, while the odd-elastic modulus here refers to microscopic odd-elastic interactions \cite{scheibner2020odd,fruchart2023odd}.


To extract the non-reciprocal interactions encoded in the kernel in Eqs. \eqref{eq:I02a2} and \eqref{eq:I02b}, we introduce a new physical quantity. Let us first consider the two-dimensional Fourier transform of the non-local elastic modulus in a space spanned by $s$ and $s'$. The transformation of ${\bl \bar{\kappa}}(s, s')$ with a wave vector {\bl $(\nu_{s}, \nu_{s'})$} is then given by
\begin{equation}
\tilde{\kappa}{\bl (\nu_{s}, \nu_{s'})}=\int\int {\bl \bar{\kappa}}(s,s')e^{-i(\nu_{s}s+\nu_{s'}s')}\,dsds'
\label{eq:I21a},
\end{equation}
where the integral is performed over the two Lagrangian coordinates, {\bl i.e., $(s,s')\in[0,\ell]\times[0,\ell]$}. 

{\bl To characterize the interactions between the units of material, it is  useful to decompose the wave vector into diagonal and perpendicular components (See Fig. \ref{fig:mat-schmat}), rather than horizontal and vertical components. The diagonal part of the elastic matrix, along the $\nu$-direction in Fig. \ref{fig:mat-schmat}, represents spatial variations of the ordinary elastic response, while the off-diagonal parts represent non-local interactions. To characterize this non-local behavior, we consider the wave vector along the $\hat{\nu}$-direction in Fig. \ref{fig:mat-schmat}, which is perpendicular to the diagonal direction, given by $\nu_{s}+\nu_{s'}=0$.} 

By plugging this relation into Eq. \eqref{eq:I21a} and introducing $\hat{\nu}=(\nu_{s}-\nu_{s'})/2$, we can obtain a complex function,
\begin{equation}
\tilde{\kappa}(\hat{\nu})=\int\int {\bl \bar{\kappa}}(s,s')e^{-i\hat{\nu}(s-s')}\,dsds'
\label{eq:I21b},
\end{equation}
which we hereafter call the {\it odd-elastic modulus}. {\bl The physical meaning of this quantity becomes clearer when the kernel function is decomposed into reciprocal and non-reciprocal components, which are symmetric and anti-symmetric, respectively, with respect to the exchange of two material points $(s,s')\mapsto (s',s)$ as
\begin{equation}
\bar{\kappa}_{\textrm{e}}(s,s')=\bar{\kappa}_{\textrm{e}}(s',s) ~\textrm{and}~ 
\bar{\kappa}_{\textrm{o}}(s,s')=-\bar{\kappa}_{\textrm{o}}(s',s)    
\label{eq:I21b2}.
\end{equation}
Substituting this decomposition into Eq.\eqref{eq:I21b}, we have $\tilde{\kappa}(\hat{\nu})=\tilde{\kappa}_{\textrm{e}}(\hat{\nu})+\tilde{\kappa}_{\textrm{o}}(\hat{\nu})$ with
\begin{eqnarray}
\tilde{\kappa}_{\textrm{e}}&=&
\int\int \bar{\kappa}_{\textrm{e}}(s,s')e^{-i\hat{\nu}(s-s')}\,dsds' \label{eq:I21b3a}, \\ 
\tilde{\kappa}_{\textrm{o}}&=&\int\int \bar{\kappa}_{\textrm{o}}(s,s')e^{-i\hat{\nu}(s-s')}\,dsds'
    \label{eq:I21b3b}.
\end{eqnarray}
From the relations in Eq. \eqref{eq:I21b2}, we find that $\tilde{\kappa}_{\textrm{e}}$ is a real function, whereas $\tilde{\kappa}_{\textrm{o}}$  is a pure imaginary function. Hence, the newly introduced odd-elastic modulus allows us to characterize reciprocal and non-reciprocal interactions using their real and imaginary components.
}


The odd-elastic modulus  $\tilde{\kappa}$ is equivalent to the Fourier spectrum if the elastic interactions only depend on $s-s'$. 
By definition, it is readily found that the real part of the odd-elastic modulus is an even function of $\hat{\nu}$, and this represents the even elasticity, characterizing the non-local, reciprocal elastic interactions. The imaginary part, in contrast, encodes the odd elasticity and non-local, non-reciprocal elastic interactions, and is an odd function of $\hat{\nu}$. 

{\bl In later sections, we will examine an active elastic filament, where $f_\alpha$ in Eq. \eqref{eq:I02a} indicates the torque and $\sigma_\beta$ encodes the relative angle as in the case of the Purcell swimmer. Then, the density $\bar{\kappa}(s,s')$ represents an extension of the bending modulus, while $\bar{\sigma}(s')$ corresponds to the local curvature. In this case, we will call the odd-elastic modulus $\tilde{\kappa}$ an {\it odd-bending modulus}, because it generalizes the linear relation between the torque and local curvature of the filament. }



\section{Odd-elastic modulus for active filaments}
\label{sec:odd_modulus}

In the previous sections, we examined the properties of a general odd-elastic material around the Hopf bifurcation. In the following  sections, focusing on one-dimensional active filaments such as the flagella of {\it Chlamydomonas} and sperm cells, we will exploit this framework to analyze the intrinsic elastic interactions that result in a limit cycle oscillation in the apparent shape space. We therefore assume that the swimmer shape gait is a known function obtained, for instance, from experimental observations. 

\subsection{One-dimensional elastic sphere-spring system}

\begin{figure}[!t]
\begin{center}
\begin{overpic}[width=8.8cm]{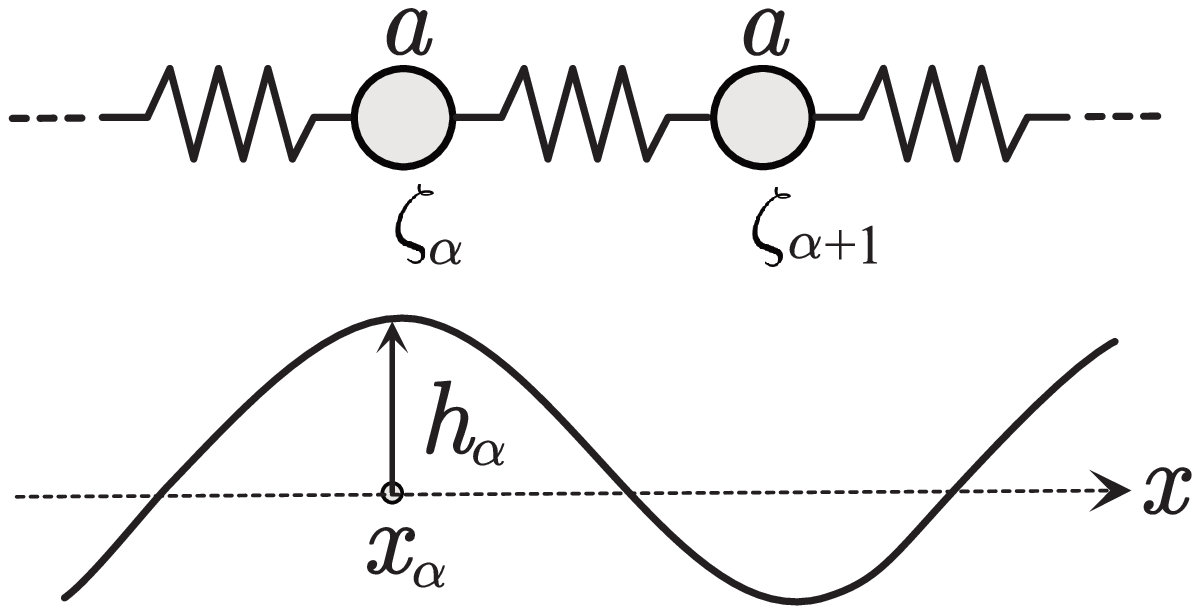}
\put(0,50){{\large (a)}}
\put(0,25){{\large (b)}}
\end{overpic}
\caption{Schematics of example models for intrinsic elasticity. (a) One-dimensional sphere-spring system. Spheres of radius $a$ are connected by linear springs
. (b) Elastic filament immersed in a viscous fluid. Assuming a small-amplitude deformation from a straight line, we parameterize the filament shape by the height $h$.}
\label{fig:solvable}
\end{center}
\end{figure}

To gain insights into the non-local, non-reciprocal interactions inside an elastic material, we start with a solvable one-dimensional model.
Our first example is a one-dimensional sphere-spring system, where $N$ spheres of radius $a$ are connected by elastic springs and form a one-dimensional array as schematically shown in Fig. \ref{fig:solvable}(a). The system is immersed in a viscous medium with viscosity $\eta$, and each sphere experiences viscous drag with a drag coefficient $\gamma=6\pi a\eta$, whereas we neglect the viscous drag on the elastic springs. We use the Lagrangian label for the sphere, $s_\alpha$, 
which corresponds to the position of a sphere at rest. We assume a periodic boundary condition, $s_{N+1}=s_1$, and the material points are equally spaced with discretization, $\Delta \ell=\ell/N$, where $\ell$ is the total length of the spatial period of the problem.

Let ${\bl \zeta}_\alpha$ be the {\bl displacement} of the sphere labeled by $\alpha$ as the shape variable $\sigma_\alpha$.
{\bl We first consider a local elastic interaction using elastic springs with a spring constant $k$ between neighboring spheres. The equation of motion is then given by}
\begin{equation}
m\ddot{{\bl \zeta}}_\alpha=-\gamma\dot{{\bl \zeta}}_\alpha+k\left( {\bl \zeta}_{\alpha+1} -2 {\bl \zeta}_{\alpha}+{\bl \zeta}_{\alpha-1}\right)
    \label{eq:I11}.
\end{equation}
{\bl 
In the large-$N$ limit, its continuous representation can be obtained, where we take the $\Delta\ell\rightarrow 0$ limit with the mass density $\bar{m}=m/\Delta \ell$, drag per unit length $\bar{\gamma}=\gamma/\Delta \ell$ and $\bar{k}=k\Delta \ell$ kept constant. This argument leads to the well-known continuum equation for the displacement field $\zeta(s, t)$ via
\begin{equation}
\bar{m}\frac{\partial^2 \zeta}{\partial t^2}=-\bar{\gamma}\frac{\partial\zeta}{\partial t}+\bar{k}\frac{\partial^2 \zeta}{\partial s^2}
    \label{eq:I11a}.
\end{equation}
}

{\bl {\rd What we examine by odd-elastic modulus takes a different approach; we specify the interactions for a given wave pattern, rather than deriving differential equations from local microscopic interactions.}}
{\bl {\bl Hence,} instead of Eq. \eqref{eq:I11}, we {\bl consider} a general non-local elastic force given by Eq. \eqref{eq:I02a}, and our {\bl non-local} sphere-spring system {\bl as}}
\begin{equation}
m\ddot{{\bl \zeta}}_\alpha=-\gamma\dot{{\bl \zeta}}_\alpha {\bl -}K_{\alpha\beta}{\bl \zeta}_\beta
    \label{eq:I12}.
\end{equation}
We then solve $K_{\alpha\beta}$ for a given wave pattern $\zeta_\alpha(s_\alpha,t)$. 

We first consider the frictionless case, where $\gamma=0$. Our goal here is to calculate the kernel function $K_{\alpha\beta}$ that sustains a given traveling wave. To do so, we consider a traveling wave pattern with wavenumber $\nu$ and angular frequency $\omega$ as 
\begin{equation}
    {\bl \zeta}_{\alpha}=A\sin(\nu s_\alpha -\omega t)
    \label{eq:I12a}
\end{equation} with an arbitrary amplitude $A$. 
Then, we set an orthogonal basis ${\bf W}=(\bm{w}^{(1)}, \bm{w}^{(2)},\cdots)$, such that $w^{(1)}_\alpha=\sin(\nu s_\alpha)$ and $w^{(2)}_\alpha=\cos(\nu s_\alpha)$.  Through direct calculations of ${\bl \zeta}_\alpha$ and $\ddot{{\bl \zeta}}_\alpha$ and comparison of both terms in Eq. \eqref{eq:I12}, 
we readily obtain 
\begin{equation}
{\bf K}={\bl m\omega^2}{\bf W}
\begin{pmatrix}
1 &    &    \\
   & 1 &   \\
   &    & {\bf O} \\
\end{pmatrix}
{\bf W}^{\textrm{T}}
\label{eq:I13c},
\end{equation}
where the symbol in the bottom right block ${\bf O}$ indicates that non-designated components are all zero.
This leads to a matrix representation of the non-local elasticity,
\begin{equation}
K_{\alpha\beta}={\bl m\omega^2}\cos[\nu(s_\alpha-s_\beta)]
\label{eq:I13a}.
\end{equation}
{\bl The continuum representation is obtained by taking the large-$N$ limit, with the mass density $\bar{m}$ (mass per unit length) kept constant, as} 
\begin{equation}
{\bl \bar{\kappa}}(s,s')={\bl \bar{m}\omega^2}\cos[\nu(s-s')]
\label{eq:I13b}.
\end{equation}
{\bl Here, the scaling $\bar{\kappa}=\kappa/\Delta \ell$ is different from that used for Eq. \eqref{eq:I11a}, where $\bar{k}=k\Delta \ell$ is assumed to be constant.}

{\bl 
The wavenumber of the propagating wave is selective in the non-local sphere-spring system, which is qualitatively different from the wave equation \eqref{eq:I12}, where the interactions are local and waves with an arbitrary frequency may propagate.
These differences become clearer when we solve the non-local sphere-spring system using the kernel representation of Eq. \eqref{eq:I13a}.

To do so, we first consider a time-frequency Fourier transform, $\zeta_\alpha(t)=\int \xi_\alpha(p)e^{-ipt}dp$, where $\xi_{\alpha}(p)$ is a Fourier component with frequency $p$. We then have 
\begin{equation}
\omega^2{\bf W}\begin{pmatrix}
1 &    &    \\
   & 1 &   \\
   &    & {\bf O} \\
\end{pmatrix}
{\bf W}^{\textrm{T}}\bm{\xi}(p)=p^2\bm{\xi}(p)
    \label{eq:I13a1},
\end{equation}
which 
forms an eigenvalue problem. This may be exactly solved and the inverse Fourier transform leads to a general solution, given by
\begin{eqnarray}
\zeta_\alpha&=&c_1\sin(\nu s_{\alpha}-\omega t+\varphi_1)
\nonumber \\ &&+
c_2\cos(\nu s_{\alpha}-\omega t+\varphi_2)
+\sum_{\alpha=3}^N c_\alpha w_\alpha^{(k)}
    \label{eq:I13a0},
\end{eqnarray}
with constants $\varphi_1$, $\varphi_2$ and $c_\alpha$ ($\alpha=1, \dots, N)$ determined by initial conditions. As the final term in Eq. \eqref{eq:I13a0} is a time-independent constant, only a sinusoidal wave with wavenumber $\nu$ can propagate, and thus the wave pattern is robust against disturbance. 
}

We then consider the same oscillatory behavior for the overdamped limit of the dynamics in Eq. \eqref{eq:I12}, that is, $m=0$. Due to viscous drag, to sustain the traveling wave, we need to inject some energy into the system. 
{\bl We therefore expect that the elastic representation of $K_{\alpha\beta}$ should contain non-reciprocal, odd components. }

Using an analysis similar to that for the frictionless dynamics above, we then have
\begin{equation}
{\bf K}={\bl \gamma\omega}{\bf W}
\begin{pmatrix}
 & 1   &    \\
 -1  &  &   \\
   &    & {\bf O} \\
\end{pmatrix}
{\bf W}^{\textrm{T}}
\label{eq:I15c},
\end{equation}
yielding the matrix representation, 
\begin{equation}
K_{\alpha\beta}={\bl \gamma\omega}\sin[\nu(s_\alpha-s_\beta)]
\label{eq:I15a}.
\end{equation}
{\bl The continuum representation is also obtained using similar arguments} as 
\begin{equation}
{\bl \bar{\kappa}}(s,s')={\bl \bar{\gamma} \omega}\sin[\nu(s-s')]
\label{eq:I15b},
\end{equation}
{\bl where the drag per unit length $\bar{\gamma}=\gamma/\Delta l$ is assumed to be constant in the large-$N$ limit.}

{\bl 
For overdamped dynamics ($m=0$), the elastic interactions  are non-local, as in the previous case, but no longer reciprocal. The matrix form, Eq. \eqref{eq:I15a}, not only selects a specific wavenumber but also sustains the associated traveling wave by the non-reciprocal components.}
Moreover, the elasticity is purely odd in the sense that the matrix is {\bl anti}-symmetric, that is, ${\bf K}=-{\bf K}^{\textrm{T}}$ and ${\bl \bar{\kappa}}(s,s')=-{\bl \bar{\kappa}}(s',s)$. 

In the underdamped case with non-zero $m$ and $\gamma$, the results are the sum of the even and odd terms. 
The necessity for energy injection is consistent with the existence of the odd elasticity in the interactions, and the injected energy is dissipated by the viscosity of the medium. 

The odd-elastic moduli for the frictionless system and the overdamped system are simply calculated from the Fourier transform of Eqs. \eqref{eq:I13b} and \eqref{eq:I15b} as
\begin{equation}
\tilde{\kappa}(\hat{\nu})=\frac{{\bl \bar{m}}\omega^2}{2}\left[ \delta(\hat{\nu}-\nu)+\delta(\hat{\nu}+\nu)\right]
\label{eq:F22a}
\end{equation}
and
\begin{equation}
\tilde{\kappa}(\hat{\nu})=\frac{i{\bl \bar{\gamma}}\omega}{2}\left[ \delta(\hat{\nu}-\nu)-\delta(\hat{\nu}+\nu)\right]
\label{eq:F22b},
\end{equation}
respectively, where {\bl $\nu$ is the wavenumber for a given wave pattern in Eq. \eqref{eq:I12a} and} $\delta(x)$ is again the Dirac delta function.
$\tilde{\kappa}$ is a real, even function in the frictionless system, while it is a purely imaginary, odd function in the overdamped system. The wavenumber of the traveling wave is clearly captured by the singular peaks.

{\bl
In summary, the sphere-spring example shows that {\bl the odd-elastic modulus {\rd captures} the internal mechanisms that robustly and selectively sustain a given wave pattern. Moreover, by considering its real and imaginary parts, we can distinguish the conservative, reciprocal elastic force from the 
 non-reciprocal force, the latter of which is associated with the energy input required in a dissipative environment.} 
}

\subsection{Small-amplitude elastohydrodynamic filament}

As a second example of the use of the intrinsic non-local elastic matrix, we examine the small-amplitude dynamics of an active filament at low Reynolds number. 
This is one of the simplest cases in elastohydrodynamics and has been studied with regard to various aspects of filament dynamics in viscous fluid for more than half a century \cite{machin1958wave, fu2008beating, bayly2015analysis, man2019morphological}. In particular, the model in this section is known to be generic near the critical point of the instability \cite{camalet2000generic}.

As shown schematically in Fig. \ref{fig:solvable}(b), we consider an infinitely long filament and let $h(x,t)$ be the height of the elastic filament from the $x$-axis, with its projection to the $x$-axis. We impose a periodic boundary condition with a spatial period of $\ell$ and discretize it by $N$ equally spaced points as in the previous example. The Lagrange label for the filament is taken by a projection onto the $x$-axis so that we have $s_\alpha=x_\alpha$. The shape variables are therefore the local height, $\sigma_\alpha=h(x_\alpha)$. Under the assumption of a small-amplitude oscillation, the continuous limit of the elastohydrodynamics of the filament follows a partial differential equation given by \cite{camalet2000generic, gadelha2010nonlinear}
\begin{equation}
\xi_{\perp}\frac{\partial h}{\partial t}=-\kappa\frac{\partial^4 h}{\partial s^4}+\frac{\partial f}{\partial s}
\label{eq:I17},
\end{equation}
where $\xi_{\perp}$ is the perpendicular drag coefficient {\bl per unit length} for a slender filament, $\kappa$ is the bending modulus, and $f(s,t)$ is the pairwise force acting on the filament {\bl per unit length} . 

In elastohydrodynamics studies, the driving force $f(s,t)$ is often a given function; otherwise, finer models for molecular activity are required to describe the dynamics of $f(s,t)$. In contrast, here, we remove the driving force term from Eq. \eqref{eq:I17} and consider instead a non-local, non-reciprocal bending modulus, via 
\begin{equation}
\xi_{\perp}\frac{\partial h}{\partial t}=-\int_{-\infty}^{\infty} {\bl \bar{\kappa}}(s, s')\frac{\partial^4 h}{\partial s^4}(s')\,ds'
\label{eq:I18},
\end{equation}
that will effectively play the role of driving actuation. We then again examine the kernel function ${\bl \bar{\kappa}}(s, s')$ that produces a  sinusoidal traveling wave $h(s,t)=A \sin(\nu s-\omega t)$.
To sustain this wave,  through calculations similar to those given above, we obtain the non-local elastic kernel as
\begin{equation}
{\bl \bar{\kappa}}(s,s')=\frac{\xi_\perp \omega}{\nu^4}\sin[\nu (s-s')]
\label{eq:I19}.
\end{equation}  
As in the overdamped case of the sphere-spring system, this is non-reciprocal and purely odd in terms of the exchange of the positions, while no even components emerge.


{\bl Although we use a continuum equation in this example, we can start from its discrete version, where the relative angles between neighboring angles are used as shape parameters. This formulation is used in the next section to deal with more general elastohydrodynamic interactions, such as finite-amplitude flagellar waveforms, where we perform numerical estimations.
}






\section{Odd-elastic modulus for biological swimmers}
\label{sec:flag_modulus}

\begin{figure}[!t]
\begin{center}
\begin{overpic}[width=8.5cm]{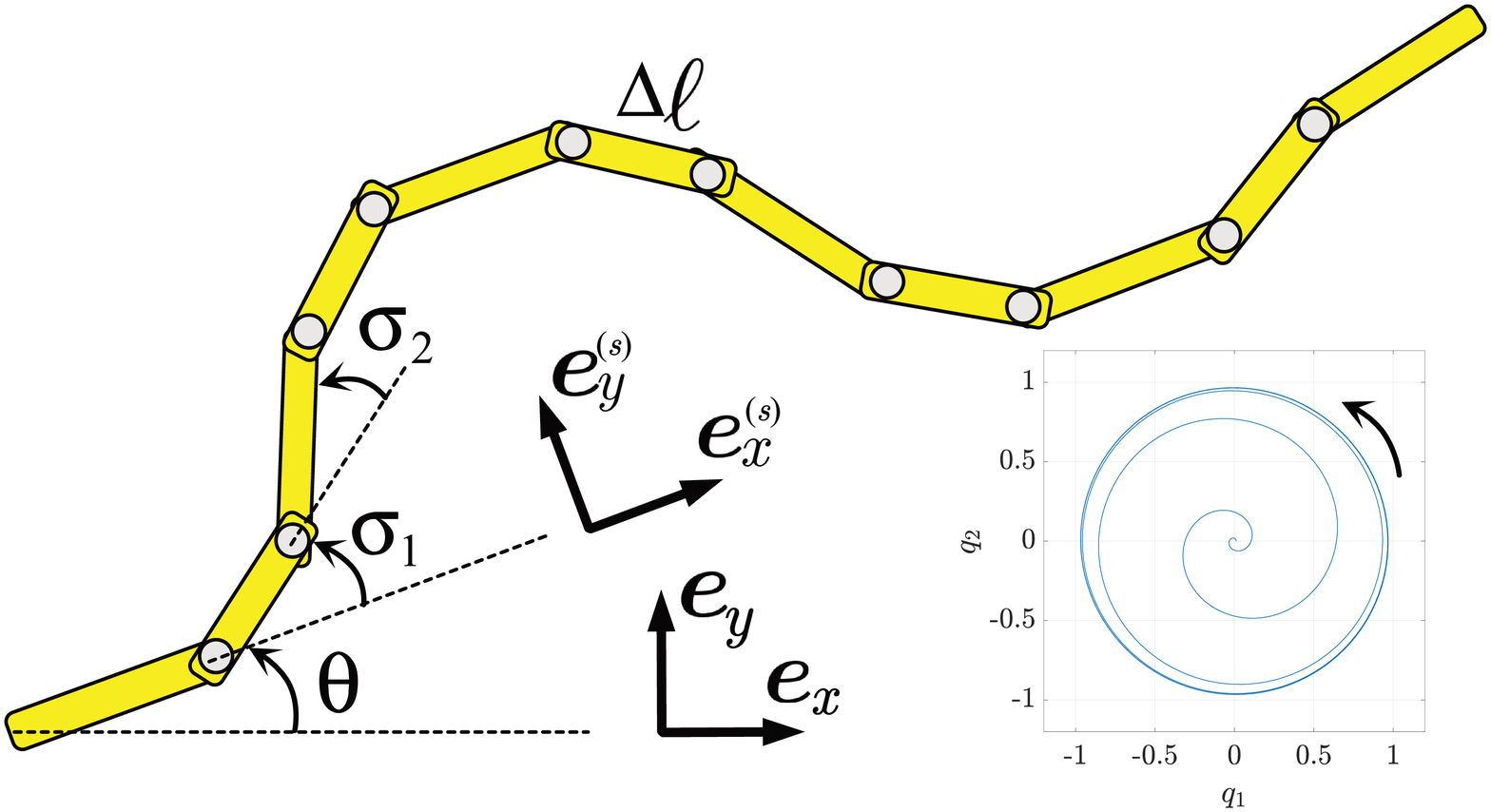}
\end{overpic}
\caption{Schematic of coarse-grained representation of elastic filament and typical dynamics in $q_1-q_2$ shape space. We represent the flagellum by $N+1$ rods of length $\Delta \ell$, which are connected at each end by elastic hinges. The shape configuration is then specified by the relative angles between neighboring rods, denoted by $\sigma_\alpha$ ($\alpha=1, 2, \cdots, N$). The rotation of the filament is described by the angle between the horizontal axis and the first rod denoted by $\theta$. [Inset] The shape gait is given by the autonomous system and we show its typical trajectory in the $q_1-q_2$ shape space. The flagellar waveform approaches a periodic pattern described by the stable limit cycle, after starting from the initial point near the origin.
}
\label{fig:config}
\end{center}
\end{figure}

\begin{figure*}[!tb]
\begin{center}
\vspace*{0.5cm}
\begin{overpic}
[width=4.3cm]{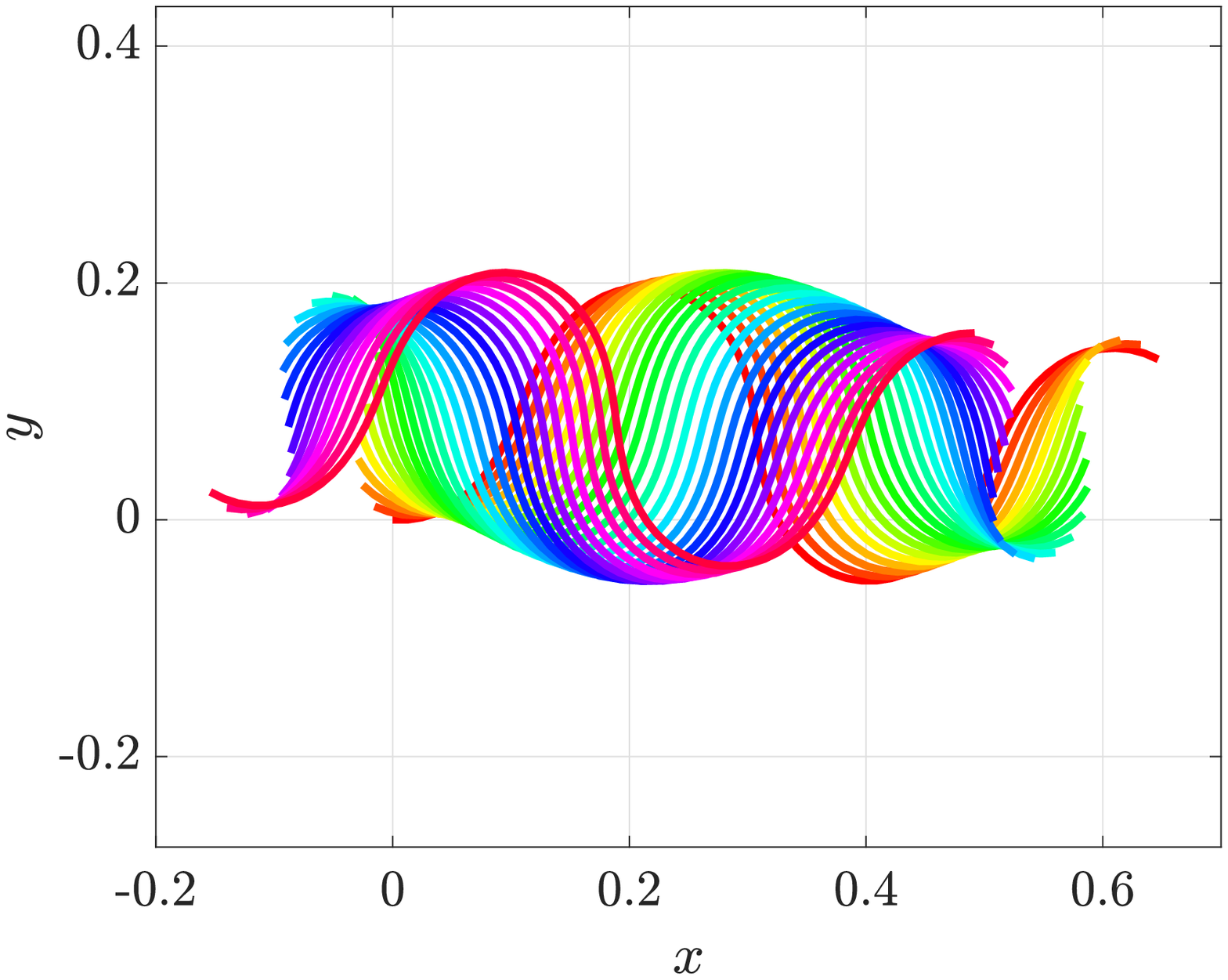}
\put(-5,75){(a)}
\end{overpic}
~\begin{overpic}
[width=4.3cm]{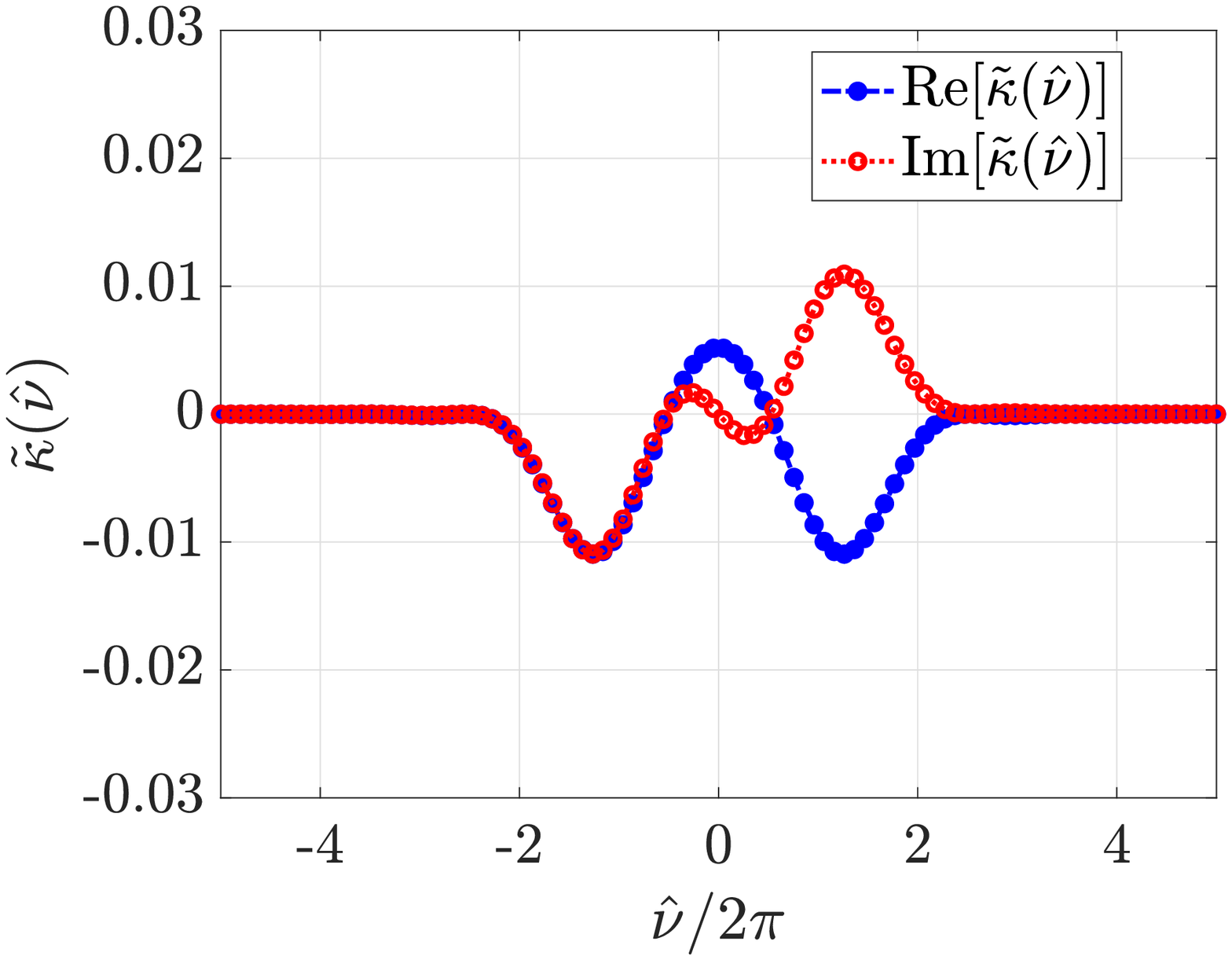}
\put(-3,75){(b)}
\put(47,80){$r=0$}
\end{overpic}
~\begin{overpic}
[width=4.3cm]{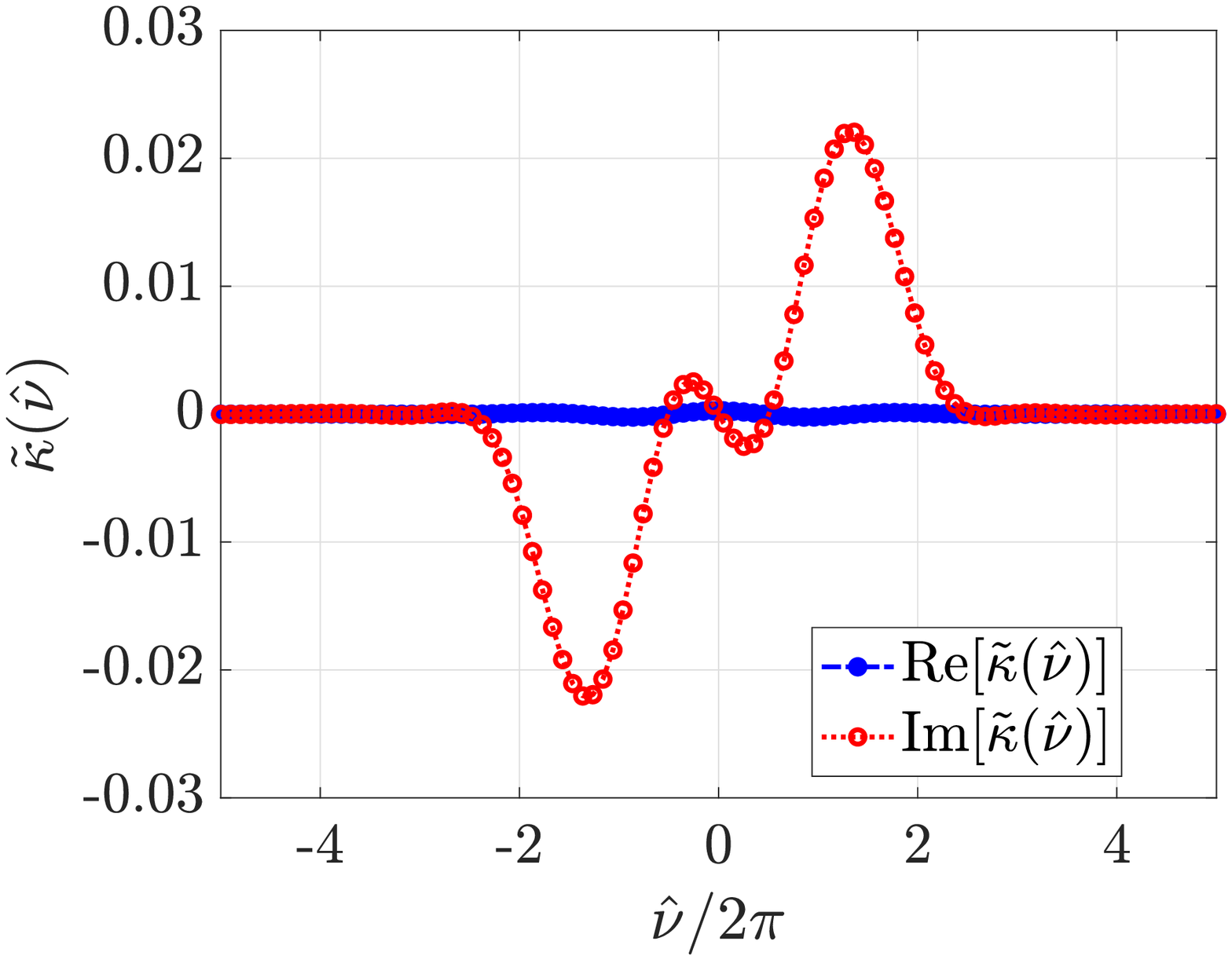}
\put(-3,75){(c)}
\put(47,80){$r=1$}
\end{overpic}
~\begin{overpic}
[width=4.3cm]{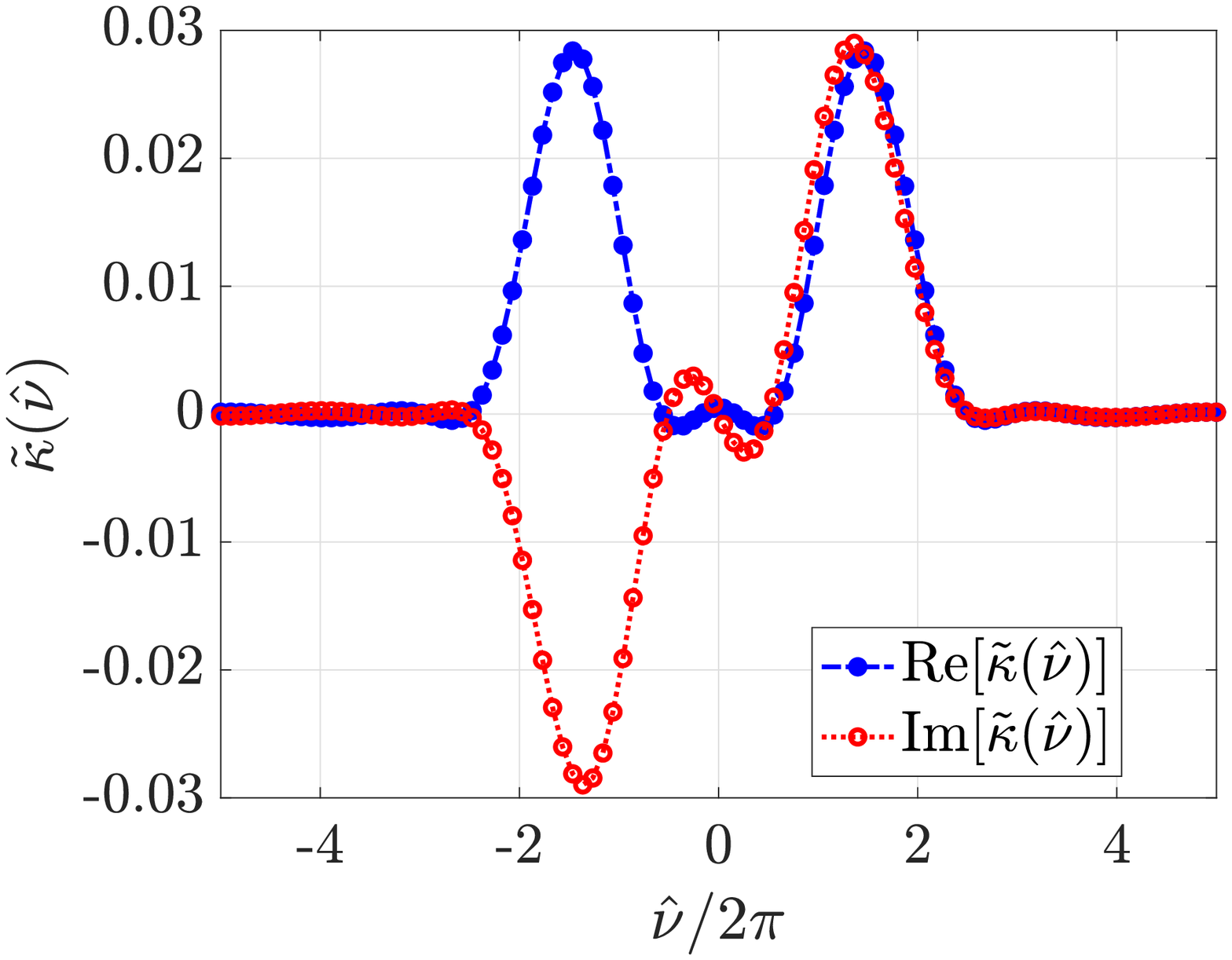}
\put(-3,75){(d)}
\put(46,80){$r=1.5$}
\end{overpic}
\caption{Flagellar waveforms and odd-bending modulus along circle with radius $r=\sqrt{q_1^2+q_2^2}$ in two-dimensional shape space for sinusoidal flagellum with wavenumber $\nu/2\pi=1.5$: (a) superposed waveforms for a swimming flagellum during one beat cycle with its left end initially located at $(x, y)=(0, 0)$, (b) $r=0$ at the origin, (c) $r=1$ on the limit cycle orbit, and (d) $r=1.5$ outside the limit cycle. In each panel, the real and imaginary parts are shown by the blue broken curve and the red dotted curve, respectively. The stable limit cycle corresponds to a circle of radius $r=1$. }
\label{fig:sin}
\end{center}
\end{figure*}


In this section, to further clarify intrinsic non-local interactions in flagellar swimming, we investigate finite-amplitude biological models of flagellar waveforms in a representative sperm cell and {\it Chlamydomonas} beat, in addition to experimental data for human sperm flagella provided by Ishimoto et al. \cite{ishimoto2018human}.

The elastic properties of a flagellum are often modeled as an Euler-Bernoulli beam, in which the local elastic moments are proportional to the local curvature of the filament. To generate the flagellum wave, as seen in the small-amplitude example, we require internal actuation in the flagellar model. Our theory of odd elastohydrodynamics developed in Section \ref{sec:setup}, however, incorporates the internal actuation as non-local generalized elasticity, together with the passive elastic response. In this section,  we numerically examine the intrinsic elasticity of a swimming filament by analyzing the given limit cycle dynamics in shape space (see the inset of Fig. \ref{fig:config}).

\subsection{Numerical methods}

To implement the flagellar elastohydrodynamics, we model the swimming flagellum as a slender filament moving in a plane and represent its dynamics by the linkages model \cite{moreau2018asymptotic}, where $N+1$ rods of length $\Delta\ell$ are connected at each end by elastic hinges to form a single filament. The shape configuration is then specified by the relative angles between neighboring rods, denoted by $\sigma_\alpha$ ($\alpha=1, 2, \cdots, N$; Fig. \ref{fig:config}). These relative angles are discretized representations of the local curvature, and for a passive elastic filament, a linear elastic torque is applied at each hinge. Here, however, we generalize this torque to non-local, nonlinear interactions, analogous to the dynamics of Eq. \eqref{eq:N10a}.

The shape gait of the filament is described by a stable limit cycle in the $q_1-q_2$ apparent shape space (inset of Fig. \ref{fig:config}).
To transform the two representations between $\bm{\sigma}$ and $\bm{q}$, let us introduce $\bm{w}^{(\alpha)}$ as the lowest $N$ PCA modes obtained 
either from data generated through a mathematical model or experimental data.
The number of intrinsic shape coordinates should be smaller than the PCA modes $N_{\textrm{flag}}$ in the original data. We then expand the intrinsic shape coordinates in the PCA modes $\bm{w}^{(\alpha)}$ as $\bm{\sigma}=q_\alpha\bm{w}^{(\alpha)}$ or  $\bm{\sigma}={\bf W}\bm{q}$ in a matrix form, where ${\bf W}=(\bm{w}^{(1)}, \bm{w}^{(2)}, \cdots, \bm{w}^{(N)})\in \textrm{O}(N)$ form an orthonormal basis of the shape space. To apply our theory, 
we consider the flagellar waveforms described by Eqs. \eqref{eq:N10a}--\eqref{eq:M11b}, 
and assume the form of the damping modes appearing in the right-bottom block in Eq. \eqref{eq:N10a}, given by ${\bf \hat{K}}^{\textrm{d}}=\kd{\bf I}_{N-2}$, where $\kd$ is a non-negative constant and ${\bf I}_{N-2}$ is the $(N-2)$-dimensional identity matrix.

When the flagellar waveform possesses {\bl non-zero time-averaged curvature}, $\bm{\sigma}_0$, this part cannot be captured by the PCA modes. 
The odd-elastic representation of Eq. \eqref{eq:M01b} is therefore extended to the form
\begin{equation}
\dot{\bm{\sigma}}=-{\bf Q}{\bf W} {\bf \hat{K}}{\bf W}^{\textrm{T}}(\bm{\sigma}-\bm{\sigma}_0)
\label{eq:F06a}.
\end{equation}
Also, this change affects the factor $r^2$ in Eq. \eqref{eq:M11b} as
\begin{equation}
r^2=\left(\bm{w}^{(1)}\cdot(\bm{\sigma}-\bm{\sigma}_0)\right)^2+\left(\bm{w}^{(2)}\cdot(\bm{\sigma}-\bm{\sigma}_0)\right)^2
\label{eq:F06b}.
\end{equation}

{\bl In the following examples, we {\bl consider a freely swimming sperm flagellum in Secs. \ref{sec:sin} and \ref{sec:sperm}, and a clamped {\it Chlamydomonas} flagellum in Sec. \ref{sec:chlamy}.} We neglect the sperm head and {\it Chlamydomonas} cell body in these examples}
{\bl in order to showcase the odd-bending modulus for different wave patterns.} 
The elastohydrodynamic coupling, represented by the matrix ${\bf Q}$, is numerically computed by the coarse-grained method based on resistive force theory \cite{moreau2018asymptotic}, for which we used $N=80$ links.  



\subsection{Sinusoidal flagellum model}
\label{sec:sin}

We start with a representation used as a simple but canonical model of a sperm flagellum \cite{gong2020steering}, where the local curvature, or relative angle in the discretized model, at the arc length $s_\alpha\in [0, \ell]$ is given by a sinusoidal function in the form
\begin{equation}
\sigma_\alpha=C_1\sin(\nu s_\alpha-\omega t)
    \label{eq:F11}.
\end{equation}
Here, the constants $C_1, \nu,$ and $\omega$ are the curvature amplitude, wavenumber, and beat angular frequency, respectively.
This simple sinusoidal function  is not only theoretically useful, but is also representative of many sperm flagella of marine species \cite{shiba2008ca2+, guasto2020flagellar}. Here, as a reasonable choice to match the flagellar waveforms, we set $\nu=3\pi$. The corresponding swimming dynamics are shown in Fig. \ref{fig:sin}(a).

We then non-dimensionalize the system. {\bl We employ the flagellar length for the unit of the length scale ($\ell=1$)} 
{\bl The linear odd elasticity $\ko$ is identical to the phase velocity and hence characterizes the timescale of the beat cycle, and we set $\ko=1$. After we fix the length scale and timescale, the only remaining physical unit is the force scale. We use the elastic force as the unit for the force scale by setting $k^{\textrm{e}}=1$.} There is, therefore, one dimensionless parameter remaining in the system that characterizes the ratio between the timescales for the elastic and viscous responses, and it is known as the sperm number, $\mathrm{Sp}=\ell(\xi_\perp \ko/|\ke|)^{1/4}$ \cite{moreau2018asymptotic}, once we set $\kn=1$ to fix the value $C_1$ and the radius of the limit cycle. We also find from Eq. \eqref{eq:F11} that $\kno=\kd=0$.

As a typical parameter for swimming sperm flagella, we set $\textrm{Sp}=3$ and compute the odd-bending modulus obtained from the intrinsic elastic matrix. Due to the nonlinear nature of the elasticity, the odd-bending modulus generally depends on the instantaneous configuration. Nonetheless, due to the rotational symmetry of the dynamics in shape space, the odd-bending modulus is almost constant along a circle with a constant radius $r=\sqrt{q_1^2+q_2^2}$. 

In Fig. \ref{fig:sin}(b-d), we plot the odd-bending modulus at $r=0$, and its average over circles of radius $r=1$ and $1.5$, where $r=1$ corresponds to the stable limit cycle orbit.
The odd-bending modulus for the straight configuration ($r=0$) possesses a negative bending modulus in its reciprocal part around $\tilde{\nu}/2\pi=\pm 1.5$ [Fig. \ref{fig:sin}(b)], indicating that the wave pattern emerges as an instability for the straight rod. The non-reciprocal part also has a peak around $\tilde{\nu}/2\pi=\pm 1.5$, and this corresponds to the wave traveling along the flagellum.

On the limit cycle orbit [Fig. \ref{fig:sin}(c)], remarkably, the real part of the odd-bending modulus almost vanishes and the non-reciprocal interactions are dominant, as observed in the overdamped sphere-spring system described in Section \ref{sec:odd_modulus}. 
When the orbit moves outside the limit cycle, as in Fig. \ref{fig:sin}(d), due to the nonlinearity, the even elasticity is enhanced for the wavenumber $\tilde{\nu}/2\pi\approx \pm 1.5$, while the non-reciprocal interactions that cause the flagellar wave to propagate are also strengthened. 

\subsection{Chlamydomonas flagellum model}
\label{sec:chlamy}

\begin{figure}[!t]
\begin{center}
\begin{overpic}
[width=4.0cm]{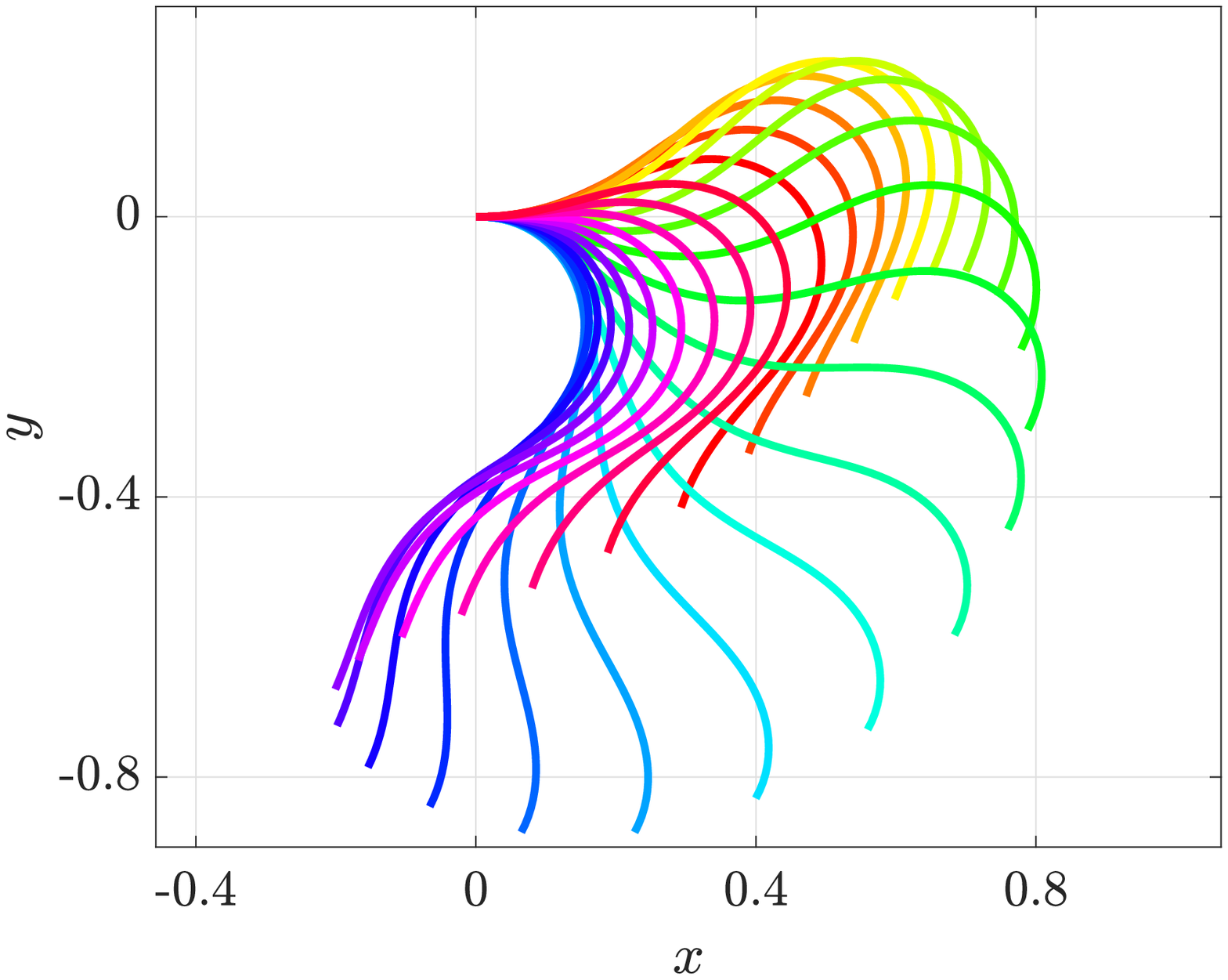}
\put(-2,76){(a)}
\end{overpic}~
\begin{overpic}
[width=4.2cm]{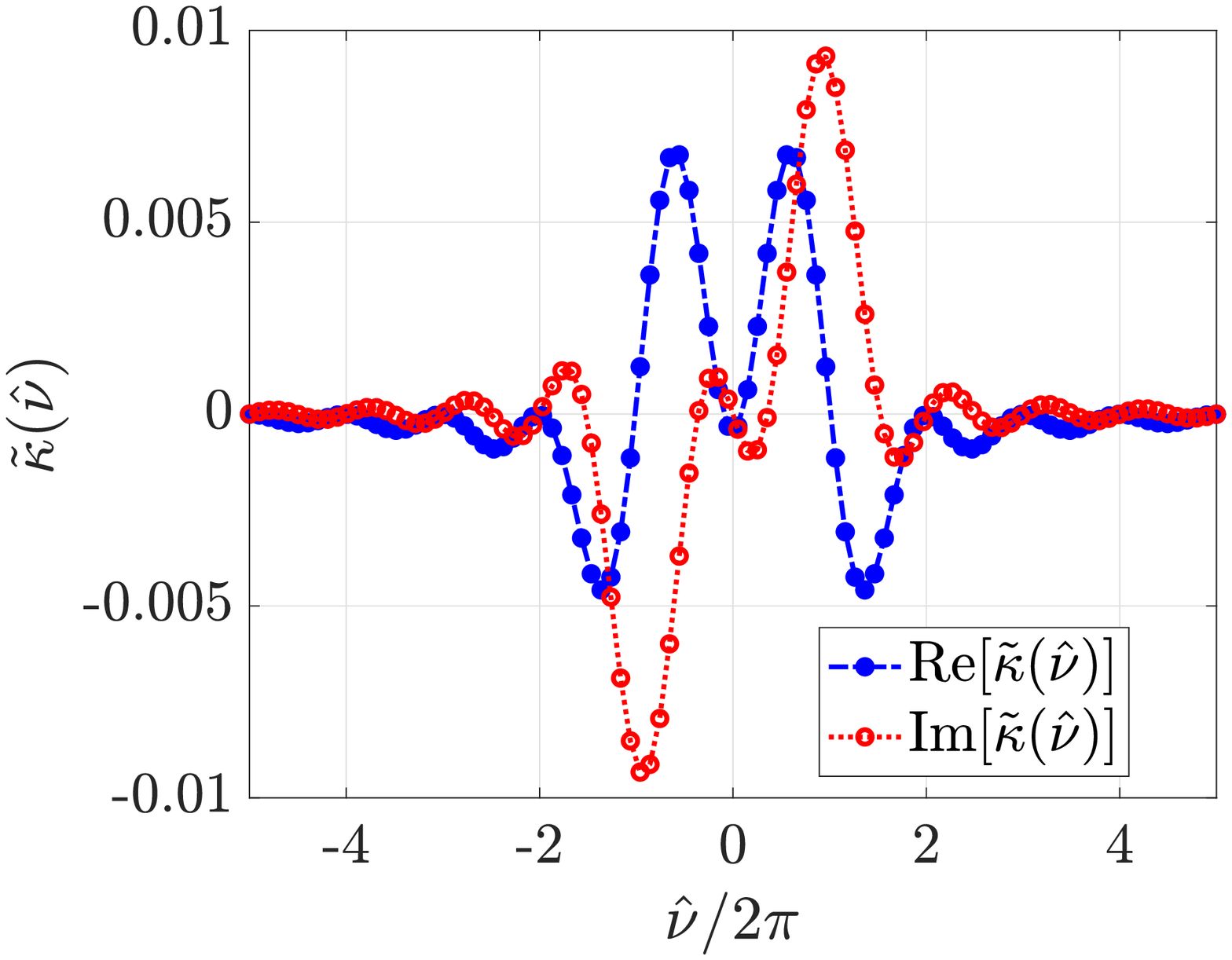}
\put(-4,72){(b)}
\end{overpic}
\caption{Flagellar waveforms and odd-bending modulus along stable limit cycle for {\it Chlamydomonas} model with proximal end clamped at $(x, y)=(0, 0)$. (a) Superposed waveforms during one beat cycle. (b) Real and imaginary parts of the odd-bending modulus.}
\label{fig:chlamy}
\end{center}
\end{figure}

We now proceed to another type of simple model, reproducing a {\it Chlamydomonas} flagellum, which is one of the most studied flagella beat patterns and characterized by its asymmetric ciliary beat pattern. According to Geyer et al. \cite{geyer2016independent}, the local curvature of the {\it C. reinhardtii} flagellum is well represented by a simple function,
\begin{equation}
\sigma_\alpha=C_0 + C_1\sin(\nu s_\alpha-\omega t)
    \label{eq:F12},
\end{equation}
as in the previous model, but with a non-zero constant $C_0$ for the mean curvature. We set $C_0=-1/40$, $\nu=2\pi$, $\kn=25/4$, and the sperm number $\textrm{Sp}=1$ for a biologically reasonable waveform, as plotted in Fig. \ref{fig:chlamy}(a). Other parameters are the same as in Section \ref{sec:sin}.
Here, we clamped the proximal end of the  
{\it Chlamydomonas} flagellum at $(x,y)=(0,0)$ and neglected the cell body and the other flagellum for simplicity. In computing the elastohydrodynamics, we further imposed the clamped boundary condition by removing the rows for the rigid body motion in Eq. \eqref{eq:M01}.

The resulting odd-bending modulus averaged over the limit cycle is plotted in Fig. \ref{fig:chlamy}(b).
In this asymmetric beat pattern, the even part of the non-local elastic interactions remains non-zero and takes both positive and negative signs with peaks at $\hat{\nu}/2\pi\approx 0.5$ and $\hat{\nu}/2\pi\approx 1.5$, respectively. In contrast, the peak for the non-reciprocal interactions coincides with the wavenumber of the flagellar waveform, indicating that the odd elasticity drives the wave as non-conservational internal actuation.

\subsection{Human sperm flagellum model}
\label{sec:sperm}

\begin{figure}[!t]
\begin{center}
\begin{overpic}
[width=4.2cm]{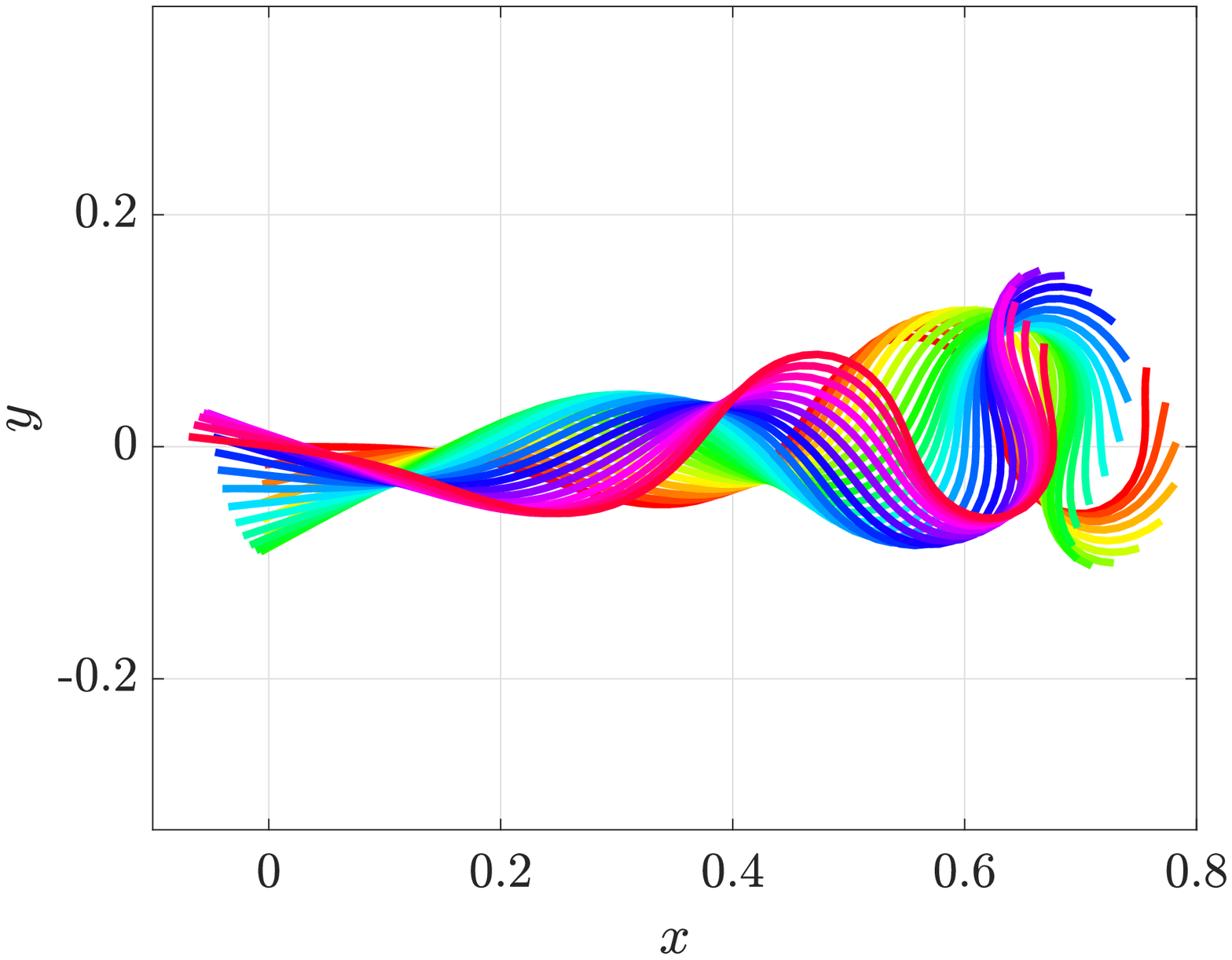}
\put(-2,73){(a)}
\end{overpic}~
\begin{overpic}
[width=4.2cm]{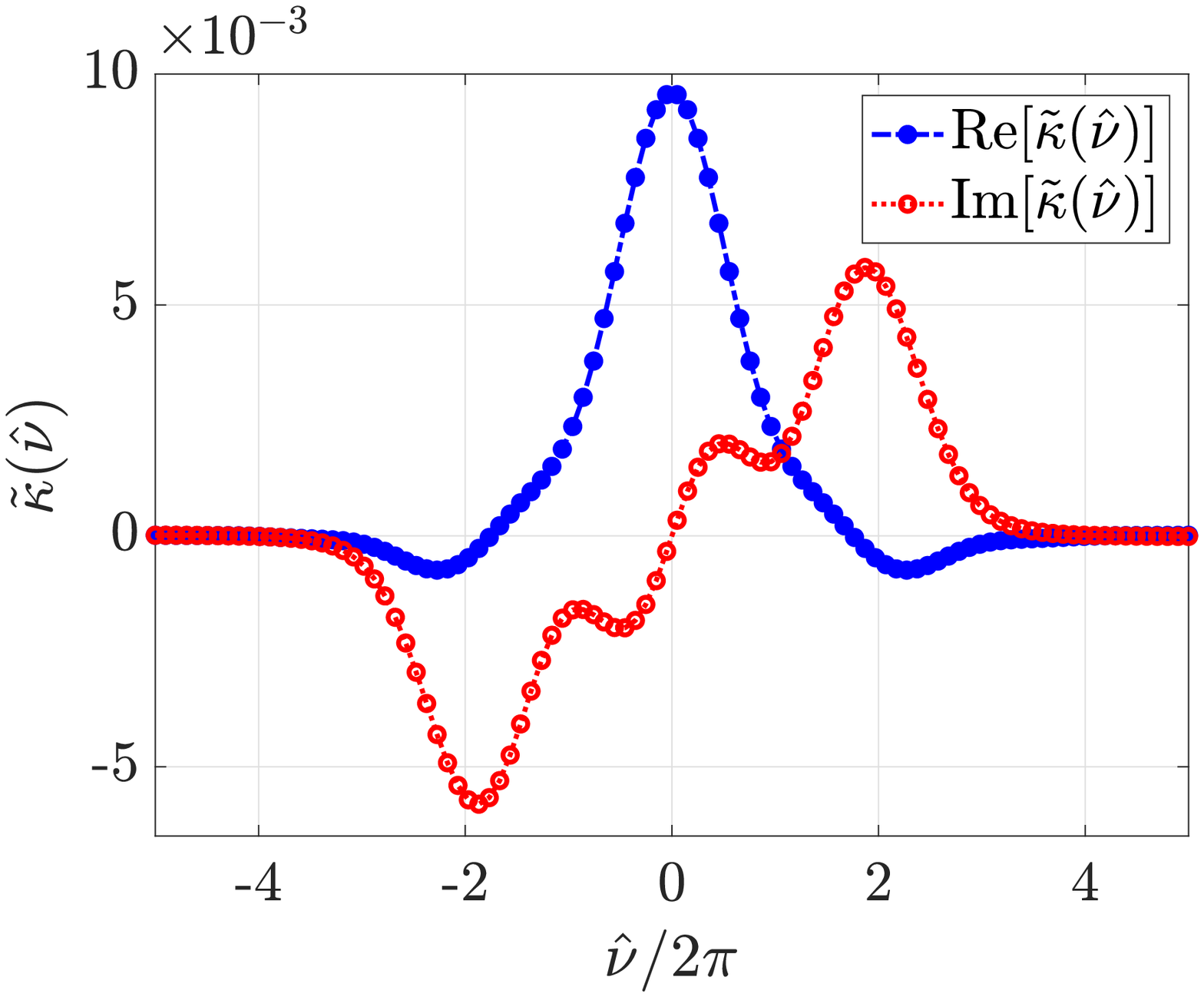}
\put(-4,72){(b)}
\end{overpic}
\caption{Flagellar waveforms and odd-bending modulus along stable limit cycle in data-driven human sperm model. (a) Superposed waveforms for a simulated sperm flagellum during one beat cycle with its left end initially located at $(x, y)=(0, 0)$. (b) Real and imaginary parts of the odd-bending modulus. }
\label{fig:spec}
\end{center}
\end{figure}

We now proceed to analyze human sperm data for our active elastic filament. 
Here, we approximate the limit cycle orbit as a unit circle in the two-dimensional flagellar PCA space using Eq. \eqref{eq:N10a} by expanding the shape variable using the flagellar PCA modes. The parameters used in this section are the same as those used in Section \ref{sec:sin}, except that we use a non-zero value of $\kd=0.1$ to ensure the existence of a stable limit cycle in the $N$-dimensional shape space. The free swimming behavior is shown in Fig. \ref{fig:spec}(a) as superposed snapshots of the flagellum.

As in the previous examples, we plot the odd-bending modulus averaged over the limit cycle in Fig. \ref{fig:spec}(b). The reciprocal interactions represented as the real part of the odd-bending modulus can take both positive and negative values depending on the wavenumber $\hat{\nu}$. The peaks around $\hat{\nu}\approx 0$ indicate the local passive elastic response of the flagellum. The negative reciprocal elasticity has a peak around a wavenumber $\hat{\nu}/2\pi\approx 2$, where the peaks for the non-reciprocal elastic interactions are also located. This are similar to the generation of the wave as a Hopf instability, as shown in Fig. \ref{fig:sin}(b). These observations, therefore, imply the following mechanical balance. The internal activity, shown by the peak in the odd elasticity and negative even elasticity, generates a flagellar wave, which is relaxed by the passive elasticity characterized by a local even elastic response.

\section{Discussion and conclusions}
\label{sec:conc}

In this study, to formulate the dynamics of a living material in a viscous fluid, we investigated a general description of swimming under a periodic limit cycle oscillation by extending the concept of odd elasticity to a nonlinear regime. By means of a change of basis from intrinsic to apparent shape coordinates, we reduced the shape dynamics to the normal form of a Hopf bifurcation, which is in turn mapped to nonlinear odd elasticity. This formulation, which we refer to as {\it odd elastohydrodynamics}, then enables us to access the internal non-local, non-reciprocal interactions in the intrinsic shape space. Further, to characterize the internal activity as well as the passive elastic response, we introduced a new concept, the {\it odd-elastic modulus}, defined by a spatial Fourier transform in an extended space. 
Of note, this odd-elastic modulus is distinct from the widely used complex modulus or dynamic modulus defined in frequency Fourier space \cite{doi2013soft}. 

With the help of the autonomous odd-elastic dynamics of an active system,
{\bl 
we were able to examine the general aspects of microswimmer dynamics. Furthermore, in the Appendices, we 
} examined in detail the effects of noise from the swimming gait on the swimming performance {\bl by extending the well-known swimming formula that provides the average swimming velocity to a general noisy limit cycle}. By calculating the probabilistic areal velocity in shape space, we found that the effect of noise on the odd elasticity is negligibly small in a small-deformation regime. 
Further analysis of the noisy limit cycle in the shape space allowed us to bridge the entropy production and work done by the odd elasticity in the nonlinear regime, and making it consistent with the physical interpretation that the internal actuation inside the elastic material is described by odd elasticity.

Then, we applied our theory to the analysis of the internal interactions of living organisms, focusing on flagellar swimming. From solvable simple models to biological flagellar waveforms for {\it Chlamydomonas} and sperm cells, we studied the odd-bending modulus to decipher the non-local, non-reciprocal inner interactions within the material.
In particular, we found that the swimmers can possess negative reciprocal even elasticity at some spatial frequencies, indicating mechanical instability by internal actuation. The imaginary part of the odd-bending modulus is the material non-reciprocal response and corresponds to the odd elasticity, which represents the speed of the generated flagellar wave. 

For the limit cycle, we found that the even elasticity ceases for some simple models, suggesting a simple nonlinear description of the material. To illustrate its usefulness in a biological context, we further analyzed the intrinsic odd-elastic response by using {\it Chlamydomonas} and human sperm flagella models, deciphering non-local elastic interactions in biological flagella.  

It is useful to point out that the current description of active elastic material includes elastohydrodynamic coupling with the outer viscous environment. We have in turn expanded the notion of odd elasticity as a stress-strain linear relation to an effective material constitutive relation that deals with the activity, elasticity, and fluid dynamics.
{\bl In this paper, we have assumed a circular trajectory in the apparent shape space as a limit cycle. However, by definition, the odd-elastic modulus can be calculated for any closed loop, indicating  potential applicability to a wide range of biological data.}

Not limited to a one-dimensional filament, our methodology is applicable to higher-dimensional materials such as active elastic membranes and bulk dynamics. The odd response of materials has also been examined in terms of odd viscosity and odd viscoelasticity \cite{banerjee2021active,lier2022passive,fruchart2023odd}, and {\bl natural extensions of the current methods to these odd materials may be expressed as a viscoelastic force representation, 
$\bm{f}=-{\bf K}\bm{\sigma}-{\bf J}\bm{\dot{\sigma}}
$, where odd viscosity is encoded by a non-symmetric matrix ${\bf J}$ coupled to the rate of deformation. Extension to an active elastic matrix in a viscoelastic medium is also an interesting future direction, where it is necessary to numerically calculate the hydrodynamic force ${\bf Q}$ from the viscoelastic fluid equation.}

The current methodology is also applicable to wet active matter systems, as these extended descriptions of active materials could be useful for simplifying the modeling of elastohydrodynamic interactions between cells. These modeling methods will therefore contribute to a better understanding of the underlying principles of collective behavior, in particular when elastohydrodynamics play an essential role, as reported for sperm population dynamics \cite{schoeller2020collective, taketoshi2023self}. Furthermore, microswimmers change their swimming pattern in response to the external environment, such as mammalian spermatozoa before and after capacitation \cite{ishimoto2016mechanical, simons2018sperm}, marine sperm cells in chemoattractant gradients \cite{shiba2008ca2+, lange2021sperm}, phototactic algae such as {\it Chlamydomonas} and {\it Volvox} under a light source\cite{bennett2015steering, de2020motility}, and ciliates in response to mechanical stimuli \cite{kunita2016ciliate, ohmura2018simple, echigoya2022switching}. Our description of odd elasticity will therefore enable unified comparisons for such diverse waveform morphologies of microswimmers to characterize the differences among species and individual cells.

\section*{Acknowledgments}
K.I. acknowledges the Japan Society for the Promotion of Science (JSPS) KAKENHI for Transformative Research Areas A (Grant No. 21H05309), and the Japan Science and Technology Agency (JST), FOREST (Grant No. JPMJFR212N). C.M. is a JSPS International Research Fellow (PE22023) and acknowledges funding support by JSPS (Grant No. 22KF0197).  K.Y. acknowledges support by a JSPS Grant-in-Aid for JSPS Fellows (Grant No. 22KJ1640). K.I., C.M., and K.Y. were supported in part by the Research Institute for Mathematical Sciences, an International Joint Usage/Research Center located at Kyoto University. \\


\begin{appendix}


\section{Swimming with noisy limit cycle}
\label{sec:swimming}

{\bl In this Appendix, to complete the general theory of odd elastohydrodynamics in the presence of noise caused by internal actuation, 
we first extend the swimming formula for the average swimming velocity to a temporally fluctuating swimming gait,  following biological observations of a noisy limit cycle in shape space. 
Using the gauge-field formulation for microswimming, we investigate the effects of internal active noise on swimming velocity. The role of odd elasticity is then further discussed in terms of non-equilibrium thermodynamics. } 



\subsection{Swimming with probability current}
\label{sec:noisy}

We {\bl now} consider the swimming formula in Eq. \eqref{eq:M05} in a statistical sense \cite{ishimoto2022self}. With a bracket indicating an ensemble average, the average swimming formula becomes
\begin{equation}
\langle \mathcal{A}\rangle=\frac{1}{2}F_{ij\alpha\beta}\langle q_\alpha\dot{q}_\beta\rangle=\textrm{Tr}(\mathcal{F}{\bf J})
\label{eq:N27a},
\end{equation}
where the trace is taken over the shape components. The {\bl anti}-symmetric matrix ${\bf J}$ is the probabilistic areal velocity matrix, given by
\begin{equation}
J_{\alpha\beta}=\bigg\langle \oint q_\alpha\,dq_\beta\bigg\rangle
\label{eq:N27b}.
\end{equation}
In the two-dimensional shape space, this statistical swimming formula \eqref{eq:N27a} simply reads as
\begin{equation}
\langle A_{ij}\rangle=2F_{ij12}J
\label{eq:N27b}
\end{equation}
if we write $J_{\alpha\beta}=J\epsilon_{\alpha\beta}$. The swimming speed in the form of a gauge potential is proportional to the probabilistic areal velocity $J$. 

To examine the effects of a noisy shape gait on the swimming velocity, we therefore need to evaluate the value of ${\bf J}$ by introducing stochastic dynamics in shape space.


We consider an $N$-dimensional autonomous system in the apparent shape space \eqref{eq:N10a} with Gaussian white noise, given by stochastic differential equations (SDE) in the sense of Stratonovich in the form
\begin{equation}
\frac{d q_\alpha}{dt}= f_{\alpha}(\bm{q})+G_{\alpha\beta}(\bm{q})\zeta_\beta(t)
\label{eq:A01a},
\end{equation}
where the noise has a zero-mean normal Gaussian form, that is, $\langle \zeta_i(t)\rangle=0$ and $\langle \zeta_\alpha(t) \zeta_\beta(0)\rangle =\delta_{\alpha\beta}\delta(t)$ with $\delta(t)$ denoting the Dirac delta function. 
The function $f_\alpha$ corresponds to the generalized force and torque in the apparent shape coordinates and is provided by Eq. \eqref{eq:N10a} as $f_\alpha=-\hat{K}_{\alpha\beta}q_{\beta}$. The diffusion tensor is introduced as ${\bf D}=(1/2){\bf G}{\bf G}^{\textrm {T}}$ and is symmetric and positive-definite by definition. 
The corresponding Fokker-Planck equation 
for the probability distribution function $P(\bm{q}, t)$ and the probability current $\bm{j}(\bm{q}, t)$ is 
\begin{equation}
\frac{\partial P}{\partial t}+\nabla\cdot\bm{j}=0
    \label{eq:A01b2},
\end{equation}
which is explicitly given in the sense of Stratonovich by
\begin{equation}
\frac{\partial P}{\partial t}=
-\frac{\partial }{\partial q_\alpha}\left[ f_\alpha P-\frac{1}{2}\left(G_{\alpha\gamma}\frac{\partial}{\partial q_\beta}\left( G_{\beta\gamma}P\right) \right)\right]
\label{eq:A01b},
\end{equation}
where $\bm{j}$ is provided by the terms in the bracket on the right-hand side of Eq. \eqref{eq:A01b}.

For brevity, we here only consider the dynamics in the two-dimensional $q_1-q_2$ shape space on which the stable limit cycle is located. This simplification is equivalent to the assumption of ${\bf K}^{\textrm{d}}={\bf 0}$ in Eq. \eqref{eq:N10a}.  When we relax this assumption to include $N-2$ stable modes, the swimming formula in Eq. \eqref{eq:N27b} is calculated as the sum of contributions from other dimensions as discussed in detail in Appendix C of Ishimoto et al. \cite{ishimoto2022self}. 
We may then write the noise tensor ${\bf G}$ as
\begin{equation}
{\bf G}=
g_r(r)\sqrt{2D_r}\,\bm{e}_r\otimes\bm{e}_r+rg_\theta(r)\sqrt{2D_\theta}\,\bm{e}_\theta\otimes\bm{e}_\theta
\label{eq:A02},
\end{equation} 
with the unit bases denoted by $\bm{e}_r$ and $\bm{e}_\theta$ for the radial and angle directions, respectively. The dynamics of Eq. \eqref{eq:A01a}, represented in Cartesian apparent shape coordinates, is then reduced to a set of equations in polar coordinates, in the sense of Stratonovich, as
\begin{equation}
\frac{d r}{dt}=f_r(r)+g_r(r)\sqrt{2D_r}\zeta_r,~\frac{d \theta}{dt}=f_\theta(r)+g_\theta(r)\sqrt{2D_\theta}\,\zeta_\theta
\label{eq:A02a}.
\end{equation}
Here, the suffixes $r$ and $\theta$ indicate the radial and angle coordinates, respectively, and the zero-mean noise satisfies the relation $\langle \zeta_a(t) \zeta_b (0)\rangle =\delta_{ab}\delta(t)$ for $a, b \in \{r, \theta\}$.
We then rewrite the probability current $\bm{j}$ in polar coordinates as 
\begin{eqnarray}
\bm{j}&=&\left[ \left( f_r(r) +D_r g_r(r)g_r'(r)\right)P-D_r\frac{\partial}{\partial r}\left([g_r(r)]^2 P\right)\right]\bm{e}_r  \nonumber \\ &&+\left[ rf_\theta(r)P -D_\theta \frac{\partial}{\partial \theta}\left(r g_\theta(r)P\right)\right]\bm{e}_\theta
\label{eq:A03a1},
\end{eqnarray}
yielding the
 Fokker-Planck equation \eqref{eq:A01b} in the form
\begin{eqnarray}
\frac{\partial P}{\partial t}&=&
-\frac{\partial }{\partial r}\left[ \left( f_r +D_r g_rg_r'\right)P-D_r\frac{\partial}{\partial r}\left(g_r^2 P\right)\right] \nonumber \\
&&- f_\theta\frac{\partial P}{\partial \theta}+ D_\theta g_\theta \frac{\partial^2 P}{\partial \theta^2}
\label{eq:A03},
\end{eqnarray} 
where the prime symbol denotes the derivative with respect to $r$. {\bl From the rotational symmetry of the system \eqref{eq:A02a},} the steady distribution $P_{\textrm{st}}$ is independent of the angle coordinate. {\bl By using the boundary condition for the steady distribution satisfying a zero distribution at infinity, we obtain the following relation for the steady distribution, after once integrating over $r$:}
\begin{equation}
\left( f_r(r) +D_r g_r(r)g_r'(r)\right)P_{\textrm{st}}-D_r\frac{d}{d r}\left( \{g_r(r)\}^2 P_{\textrm{st}}\right)=0
\label{eq:A04a}.
\end{equation}
The steady distribution is then formally solved as
\begin{equation}
P_{\textrm{st}}(r)=\frac{N_{\textrm{st}}}{g_r(r)}
\exp\left[ \int^r \frac{f_r(x)}{D_r\{ g_r(x)\}^2} dx \right]
\label{eq:A04b},
\end{equation}
where $N_{\textrm{st}}$ is the normalization factor.
{\bl Note that the steady distribution is known to be unique \cite{risken1996fokker} and here it is } independent of both the angle coordinates and odd elasticity because of the form $f_r=-\ke r-\kn r^3$. 
For a steady distribution, the probability current $\bm{j}$ \eqref{eq:A03a1} possesses only angular components due to the condition \eqref{eq:A04a}, leading to the form
\begin{equation}
\bm{j}_{\textrm{st}}=r f_\theta(r)P_{\textrm{st}}(r)\bm{e}_\theta
    \label{eq:E01},
\end{equation}
with $f_\theta(r)=\ko +\kno r^2$ in our dynamics. The non-vanishing probability current characterizes the violation of the detailed balance or the non-reciprocity of the non-equilibrium system, and has been studied from several perspectives, such as irreversible circulation \cite{tomita1974irreversible}, curl flux \cite{wang2008potential, fang2019nonequilibrium}, and non-reciprocity \cite{yasuda2021nonreciprocality,kobayashi2022odd}.

To calculate the value of $J$, we use the internal noise model proposed by Ma et al. \cite{ma2014active} based on experimental observation of bull sperm. Their study found that the fluctuating sperm flagellar waveforms were well described by the SDE model with multiplicative white noise given by
\begin{equation}
 g_r=r ~~\textrm{and}~~ g_\theta=1
 \label{eq:A04c},
\end{equation}
where $D_r$ and $D_\theta$ correspond to the diffusion constant in the amplitude and angle coordinates, respectively. This noise term in Eq. \eqref{eq:A04c} is obtained from additive noise in the (linear) even and odd elastic constants $\ke$ and $\kn$ by the transformations $\ke\mapsto \ko +\sqrt{2D_r}\,\zeta_r(t)$ and $\ko\mapsto \ke +\sqrt{2D_\theta}\,\zeta_\theta(t)$ in the deterministic dynamics of Eq. \eqref{eq:N10a} \cite{graham1982hopf}. Because $\ke$ and $\ko$ represent the amplitude and phase speed of the limit cycle, respectively, the noise strengths $D_r$ and $D_\theta$   therefore correspond to the amplitude and angle diffusion.

\begin{figure}[!t]
\begin{center}
\hspace*{-0.5cm}
\begin{overpic}
[width=4.8cm]{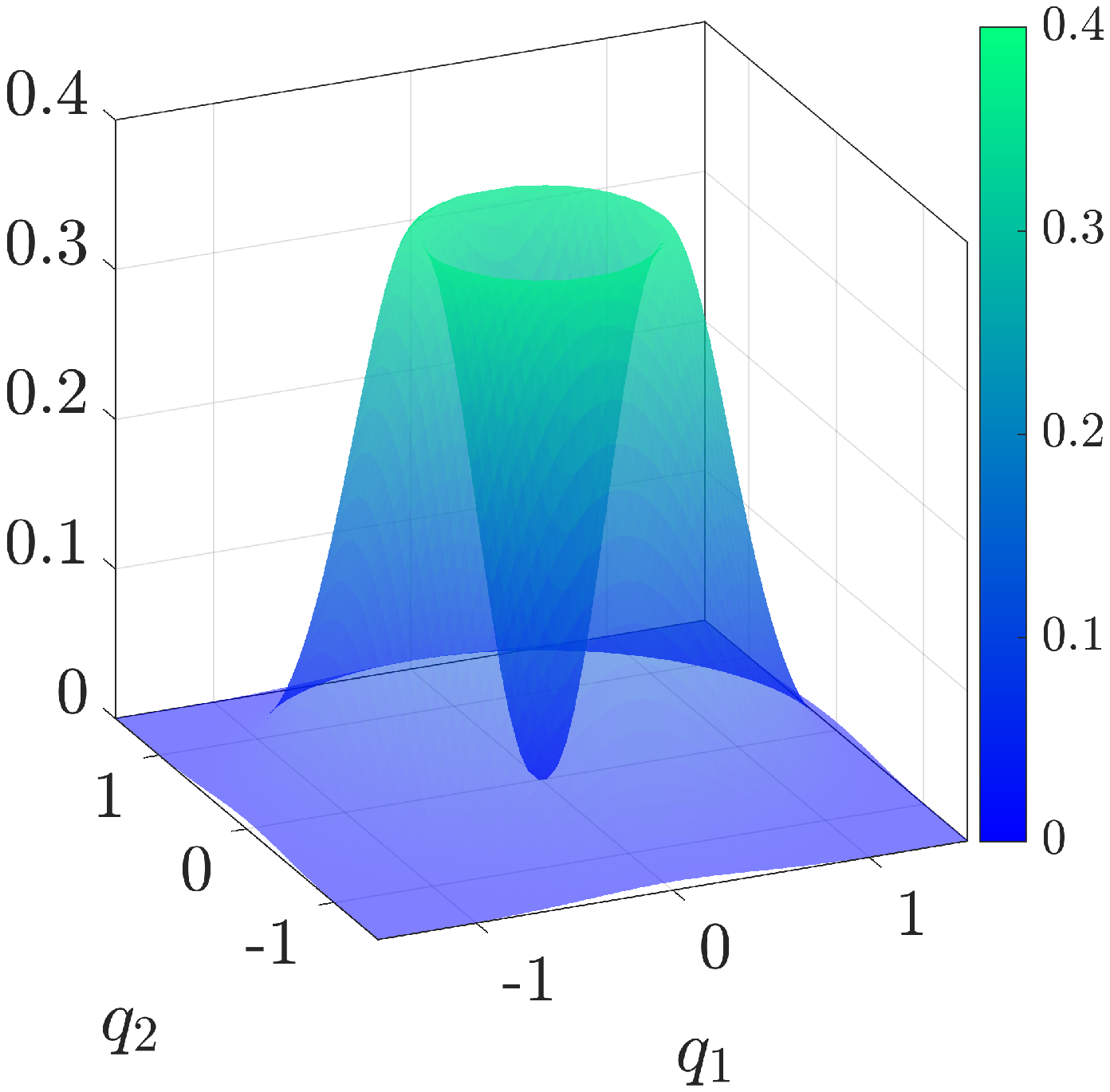}
\put(5,68){(a)}
\end{overpic}
\hspace*{-0.8cm}
\begin{overpic}
[width=4.8cm]{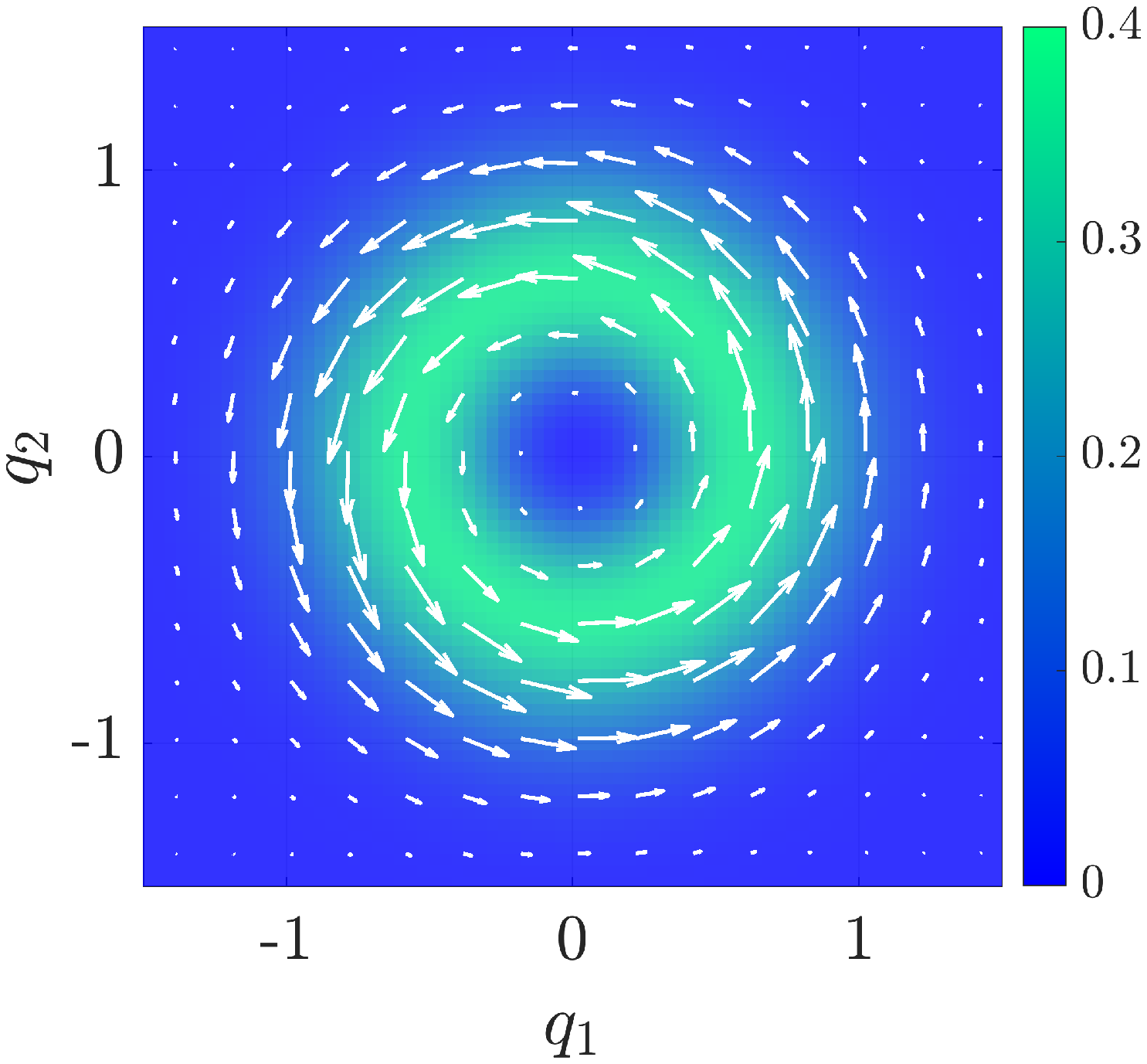}
\put(5,68){(b)}
\end{overpic}
\caption{Steady probability distribution and probabilistic current for system with noisy limit cycle with biologically relevant multiplicative noise. Parameters are set as $\ke=-1, \kn=1, \ko=1, \kno=1$, and $D_r=0.3$. (a) Steady distribution function with a volcano shape. The distribution function has a degenerate peak and reaches zero at the origin and infinity. The function is normalized so that the integrated probability becomes unity. (b) Distribution function projected onto the $q_1-q_2$ plane and superposed with probabilistic current vectors. The circular current has a maximum strength around the peak of the distribution.}
\label{fig:dist}
\end{center}
\end{figure}

The detailed form of the steady solution depends on the sign of $\ke$ \cite{graham1982hopf}. Indeed, when $\ke\geq 0$, the origin is stable and the possible steady solution is simply the Dirac delta distribution at the origin,
\begin{equation}
P_{\textrm{st}}(r)=\frac{1}{\pi}\delta(r)
\label{eq:N24a}.
\end{equation}
In the case with $\ke<0$, however, the origin is unstable and a steady limit cycle emerges, leading to 
a steady distribution in the form
\begin{equation}
P_{\textrm{st}}= N_{\textrm{st}} r^{\frac{|\ke|}{D_r}-1}\exp\left[ -\frac{\kn}{2D_r}r^2\right]
\label{eq:N24b}
\end{equation}
with a normalization prefactor $N_{\textrm{st}}$. 
This probability function vanishes at the origin when the noise is sufficiently small, $D_r<|\ke|$, whereas the distribution becomes singular at the origin once the magnitude of the noise is increased to $D_r>|\ke|$.
From the steady distribution with $|\ke|>D_r$, as plotted in Fig. \ref{fig:dist}(a), it can be seen that for the volcano-like function, the maximum values are located on a circle of radius $r^\ast=\sqrt{(|\ke|-D_r)/\kn}$, which monotonically decreases as the diffusion $D_r$ increases.

Note that the distribution in Eq. \eqref{eq:N24a} is always possible because the origin is a stationary solution of the stochastic system with our choice of noise \eqref{eq:A04c}.
The resulting steady distribution is therefore obtained by the summation of \eqref{eq:N24a} and \eqref{eq:N24b}, with their weights depending on the initial distribution. 

Let us assume $\ke<0$ and that the initial distribution does not contain the $r=0$ state. Then, the steady distribution can be further expressed with an explicit form of the normalization factor $N_{\textrm{st}}$, given by 
\begin{equation}
N_{\textrm{st}}=\frac{\sqrt{2}}{\pi}\left(\sqrt{\frac{\kn}{2D_r}} \right)^{\frac{|\ke|}{D_r}-1} \bigg/ ~\Gamma\left( \frac{|\ke|}{2D_r} \right)
\label{eq:N25b},
\end{equation}
where $\Gamma(x)$ in the denominator indicates the gamma function. Then, the steady-state probability current $\bm{j}_{\textrm{st}}$ in Eq. \ref{eq:E01} possesses a non-zero value. The plots in Fig. \ref{fig:dist}(b) show the rotational probability current superposed on the two-dimensional projection of the probability distribution. The size of $\bm{j}$ has a maximum value on a circle, which roughly overlaps with the ridge of the distribution function. Also, we can calculate $J$ as
\begin{eqnarray}
J&=&\bigg\langle \oint q_1\,dq_2\bigg\rangle
=\frac{\ko}{2}\langle r^2 \rangle+\frac{\kno}{2}\langle r^4 \rangle
\nonumber \\
&=& \frac{\ko}{2}\frac{|\ke|}{\kn}+\frac{\kno}{2}\frac{|\ke|}{\kn}\left( \frac{|\ke|}{\kn}+\frac{2D_r}{\kn}\right)
\label{eq:N26}.
\end{eqnarray}
The first term is proportional to the linear odd elasticity $\ko$ and is equivalent to the deterministic case. The noise effects only appear in the nonlinear odd-elastic term and increase (decrease) by amplifying the noise magnitude for $\kno>0$ ($\kno<0$), although the nonlinear effects are sub-dominant compared with the linear odd-elastic term. Note that this dependence is a characteristic feature of the noise form in Eq. \eqref{eq:A04c},  and the stochastic model with simple additive noise provides qualitatively similar results for a small noise case but different results in general (see Appendix \ref{app:additive} for details). 

The non-zero probability current due to the noisy limit cycle is generated only by the odd parts of the elasticity matrix, and from the swimming formula \eqref{eq:N27b}, the average swimming velocity is also  proportional to the size of the odd elasticity. 
When $\ke>0$, however, the steady probability distribution shrinks to the origin and the steady-state probability current vanishes, yielding an average swimming velocity of zero. Moreover, the noise effect does not appear in the dominant linear odd-elastic term, indicating that the fluctuating  shape gait does not impact swimming under the noisy limit cycle. However, the effects of noise depend on the stochastic model, and some qualitative differences are discussed in Appendix \ref{app:additive}.

{\bl In summary, the autonomous odd-elastic dynamics of an active system and the gauge-field formulation of microswimming enable us to examine the effects of noise due to the swimming gait on the swimming performance. Notably, we have found that the effect of noise on the odd elasticity is negligibly small in a small-deformation regime. }


\subsection{Non-reciprocity, irreversibility, and entropy production}

To conclude the general theory of odd elastohydrodynamics, we now examine the irreversible stochastic dynamics driven by the odd elasticity from the point of view of thermodynamics.
The non-zero probability current is due to the violation of the detailed balance, and this can be characterized by a non-zero (positive) entropy production rate in the non-equilibrium statistical physics \cite{evans2002fluctuation}. The averaged entropy production rate, $\dot{e}_p$, is defined by using non-equilibrium entropy $S=\int_{-\infty}^{\infty}\int_{-\infty}^{\infty} P_{\textrm{st}}\log P_{\textrm{st}}\, dq_1\,dq_2$,  the average rate of heat $\dot{Q}$, and the system temperature $T$ as
\begin{equation}
\dot{e}_p:=\dot{S}-\frac{\dot{Q}}{T}
    \label{eq:E02},
\end{equation}
which is, in general, non-negative by the thermodynamic law.
For the Langevin system given by Eq. \eqref{eq:A01a}, 
the entropy production rate is provided by \cite{feng2011potential}
\begin{equation}
\dot{e}_p= \langle v_{\alpha} D^{-1}_{\alpha\beta}v_\beta\rangle
    \label{eq:E03},
\end{equation}
where $\bm{v}$ is the probabilistic velocity for the steady state defined as $\bm{v}=\bm{j}_{\textrm{st}}/P_{\textrm{st}}$; hence, in our problem, simply $\bm{v}=r f_{\theta}(r)\bm{e}_\theta$ from Eq. \eqref{eq:E01}. By directly calculating the diffusion tensor via Eq.\eqref{eq:A02}, 
we obtain
\begin{equation}
\dot{e}_p= \frac{1}{D_\theta}\langle f^2_{\theta} \rangle 
    \label{eq:E04},
\end{equation}
which can be calculated, with $f_\theta(r)=\ko +\kno r^2$, as
\begin{equation}
\dot{e}_p= \frac{1}{D_\theta}\left[ 
(\ko)^2+2\ko\kno\frac{|\ke|}{\kn}+(\kno)^2\left( \frac{|\ke|}{\kn}+\frac{2D_r}{\kn}\right)
\right] 
    \label{eq:E05}
\end{equation}
for $\ke<0$ by using the steady-state distribution and Eqs. \eqref{eq:N24b} and \eqref{eq:N25b}. When $\ke\geq 0$, we obtain $\dot{e}_p= (\ko)^2/D_\theta$. 

These results confirm that the average entropy production rate is generated only by the odd elasticity, which agrees with the physical interpretation of the odd elasticity as an internal non-conservative activity. The average work done by the elastic force is also calculated using
\begin{equation}
\dot{W}=-\bigg\langle \oint \hat{K}_{\alpha\beta}q_{\beta} \,dq_\alpha \bigg\rangle=\langle f^2_{\theta}\rangle
    \label{eq:E06}.
\end{equation}
We note that the even elasticity is represented by a potential force and does not contribute to the average work. Comparing Eqs. \eqref{eq:E04} and \eqref{eq:E06} shows that the second law of thermodynamics reads as $\dot{e}_p=\dot{W}/k_\textrm{B}T$, with the Boltzmann constant $k_\textrm{B}$.
We thus have
\begin{equation}
D_\theta=k_\textrm{B}T
    \label{eq:E07},
\end{equation}
which is consistent with the generalized fluctuating dissipation theorem for non-equilibrium systems \cite{seifert2005entropy}.

Of note, the non-equilibrium thermodynamic relations discussed in this subsection are based on the dynamics in the shape space \eqref{eq:A01a}, or more precisely, the dynamics of Eqs. \eqref{eq:N10a} and \eqref{eq:M11b} with additional noise. The entropy production rate and the associated heat production for the full system should account for the non-reciprocal swimming motion in the physical space and the energy dissipation in the fluid. 

{\bl Nonetheless, our analysis of the noisy limit cycle in shape space bridges the entropy production and work done by the odd elasticity in the nonlinear regime, being consistent with the physical interpretation that actuation inside the elastic material is described by odd elasticity.}

\section{Probability current with simple additive noise}
\label{app:additive}

In this appendix, to complement the results of the biologically relevant multiplicative noise posited in Section \ref{sec:noisy}, we discuss the probabilistic areal velocity for the additive Gaussian noise in Eq. \eqref{eq:A02}. The form of the noise \eqref{eq:A02} is given by the setting  $G_{\alpha\beta}=\sqrt{2D}\,\delta_{\alpha\beta}$.

The formal solution \eqref{eq:A04b} then provides the steady probability distribution \cite{san1980limit}
\begin{equation}
P_{\textrm{st}}(r)=N_{\textrm{st}}\exp\left[ -\frac{\ke}{2D}r^2-\frac{\kn}{4D}r^4 \right]
\label{eq:N04},
\end{equation}
where the normalization factor $N_{\textrm{st}}$ is obtained by a direct integral of $\int_{-\infty}^\infty \int_{-\infty}^\infty P_{\textrm{st}}\,dq_1 dq_2=1$ as
\begin{equation}
N_{\textrm{st}}=\frac{1}{\pi}\exp\left( -\frac{(\ke)^2}{4D\kn} \right) \bigg/ \sqrt{\frac{\pi D}{\kn} }\,\erfc\left(\frac{\ke}{2\sqrt{D\kn}}\right)
\label{eq:N07b},
\end{equation} 
where $\textrm{erfc}(x)$ is the complementary error function.

The exponential form in Eq. \ref{eq:N04} represents the fourth-order potential associated with the nonlinear odd elasticity, and the steady distribution function is unimodal when $\ke>0$, whereas it becomes crater-shaped with a degenerate maximum on a circle with radius $r^\ast=(|\ke|/\kn)^{1/2}$ when $\ke<0$.

As in Eq. \eqref{eq:N26}, we can calculate the probabilistic areal velocity via
\begin{equation}
J=\frac{\ko}{2}\langle r^2 \rangle+\frac{\kno}{2}\langle r^4 \rangle=:J_\textrm{o}+J_\textrm{no}
\label{eq:N07c},
\end{equation}
which we separated into two terms for subsequent analyses.

After some calculations, we can express the linear odd elastic contribution, $J_\textrm{o}=(\ko/2)\langle r^2 \rangle$, as
\begin{equation}
J_\textrm{o}=
\cfrac{\ko}{2}\left(\cfrac{2\pi D}{\kn}N_{\textrm{st}}-\cfrac{\ke}{\kn}\right) 
\label{eq:N08b}.
\end{equation}
Note that $N_{\textrm{st}}$ is a non-negative function of $D$. For example, when $\ke=0$, it becomes $N_{\textrm{st}}=\sqrt{\kn/[\pi^3 D]}$. To understand the effects of noise, we need to examine the behavior of $D N_{\textrm{st}}$, and we found that this quantity monotonically increases with $D$. Indeed, by the asymptotic expression for a small $x=4D\kn/(\ke)^2>0$,
\begin{equation}
\frac{\sqrt{x}e^{-1/x}}{\textrm{erfc}(1/\sqrt{x})}=\sqrt{\pi}\left(1+\frac{x}{2}-\frac{x^2}{2}\right)+O(x^3)
\label{eq:N16a},
\end{equation}
\begin{equation}
\frac{2\pi D}{\kn}N_{\textrm{st}}-\frac{\ke}{\kn}=\frac{\ke}{2\kn}\left(x-x^2 \right)+O(x^3)
\label{eq:N16b},
\end{equation}
and we obtain an asymptotic behavior for small noise as
\begin{equation}
J_{\textrm{o}}=
\frac{\ko}{\ke} D+O(D^2) ~\textrm{when}~ \ke>0
\label{eq:N16c0}.
\end{equation}

When $\ke<0$, using the asymptotic behavior for small $x>0$, 
\begin{equation}
\frac{\sqrt{x}e^{-1/x}}{\textrm{erfc}(-1/\sqrt{x})}=\frac{1}{2}e^{-1/x}\left(\sqrt{x}+O(x^{7/2})\right)
\label{eq:N16d},
\end{equation}  
and we can obtain
\begin{equation}
J_{\textrm{o}}\simeq
\frac{\kno}{2} \frac{|\ke|}{\ke}  ~\textrm{when}~ \ke<0
\label{eq:N16c}
\end{equation}
with an exponentially small error.

We found that the average $\langle r^2\rangle$
monotonically increases with $D$, irrespective of the sign of $\ke$. 
The magnitude of the probabilistic areal velocity is therefore proportional to the odd elasticity $\ke$ and increases as the size of the noise increases, although the overall sign is determined by the sign of $\ke$. 
This noise dependence is different from that for the multiplicative noise case, although the noise effect is exponentially small for little noise.

By similar calculations, we can obtain the probabilistic areal velocity from the nonlinear odd elasticity, $J_{\textrm{no}}=(\kno/2)\langle r^4\rangle$, as
\begin{equation}
J_{\textrm{no}}=\frac{\kno}{2}\left[ \frac{2D}{\kn}-\left(\frac{\ke}{\kn}\right)\left( \frac{2\pi D}{\kn}N_{\textrm{st}}-\frac{\ke}{\kn}\right)\right]
\label{eq:N15}.
\end{equation}

By a similar asymptotic analysis, by using the expression \eqref{eq:N16a}, we obtain $J_{\textrm{no}}$ for small $D$ when $\ke>0$ as
\begin{equation}
J_{\textrm{no}}=
\frac{2\kno}{(\ke)^2} D^2+O(D^3) ~\textrm{when}~ \ke>0
\label{eq:N16c2},
\end{equation}
which is positive for small $D$ and monotonically increases (if $\kno>0$) as diffusion is enhanced. 
When $\ke<0$, again using the asymptotic \eqref{eq:N16d}, the probabilistic areal velocity is obtained for small $D$ as
\begin{equation}
J_{\textrm{no}}\simeq
\frac{\kno}{2} \left[ \frac{2D}{\kn}+\left(\frac{
\ke}{\kn}\right)^2 \right]~\textrm{when}~\ke<0
\label{eq:N16c3},
\end{equation}
with an exponentially small error.


\end{appendix}

\bibliography{library}

\end{document}